\documentclass[%
 reprint,
superscriptaddress,
 amsmath,amssymb,
aip,
jcp,
citeautoscript,
]{revtex4-2}

\usepackage{graphicx}
\usepackage{dcolumn}
\usepackage{bm}

\usepackage{xcolor}
\usepackage{float}
\usepackage{booktabs}

\begin{document}

\preprint{APS/123-QED}

\title{
The metastability of icosahedral quasicrystals assembled from one-component systems with isotropic interactions
}

\author{Cosmin A.\ Dicu-Gohoreanu}
\affiliation{Instituto de Qu\'{i}mica F\'{i}sica Blas Cabrera, Consejo Superior de Investigaciones Cient\'{i}ficas, CSIC, Calle Serrano 119, 28006 Madrid, Spain}
 \affiliation{Departamento de Química Física I (Unidad Asociada de I+D+i al CSIC), Facultad de  Ciencias Químicas, Universidad Complutense de Madrid, Madrid 28040, Spain}

\author{Jonathan P.~K.\ Doye}
\affiliation{Physical and Theoretical Chemistry Laboratory, Department of Chemistry, University of Oxford, South Parks Road, Oxford, OX1 3QZ, United Kingdom}

\author{Eva G.\ Noya}
\affiliation{Instituto de Qu\'{i}mica F\'{i}sica Blas Cabrera, Consejo Superior de Investigaciones Cient\'{i}ficas, CSIC, Calle Serrano 119, 28006 Madrid, Spain}


\date{\today}

\begin{abstract}
Icosahedral quasicrystals (IQCs) are long-range ordered materials that lack translational periodicity. While they have only been observed in multi-component metallic systems, simulations have shown that one-component IQCs can robustly self-assemble using oscillatory potentials.
However, their thermodynamic stability remains an open question. Here, we employ direct coexistence simulations to estimate the melting line of the competing solids for one such family of potentials. We find that the IQCs are never thermodynamically stable within the explored parameter space, being consistently outcompeted by periodic crystalline phases. This metastability is primarily driven by the higher internal energy of the IQC relative to the competing solids.
Our findings suggest that the observed formation of these single-component IQCs in simulations is likely a kinetic phenomenon rather than a thermodynamic one.
\end{abstract}

\maketitle

\section{\label{sec:level1}Introduction}

The discovery of quasicrystals challenged a fundamental assumption of crystallography by demonstrating that long-range translational order is not a prerequisite for sharp diffraction patterns \cite{Shechtman84}. Among these, icosahedral quasicrystals (IQCs) are of particular interest due to their high symmetry. While IQCs have only been observed in nature in metallic systems with at least two components \cite{Tsai_2000}, other quasicrystalline symmetries have been successfully identified in soft matter, in some instances in one-component systems \cite{Zeng04,Dotera11}. In contrast, computer simulations have reported that IQCs can readily self-assemble from the fluid phase in one-component models, ranging from isotropic oscillatory pair potentials (OPP) \cite{Engel2015,Damasceno2017,Dshemuchadse21,pan_dshemuchadse_23} to patchy colloids with rationally designed directional interactions \cite{Noya21,Noya25}. An open question in the field is whether these structures can be thermodynamically stable in one-component systems or if they are merely long-lived metastable states favoured by kinetic pathways during self-assembly.

The oscillatory pair potential (OPP) proposed by Engel \emph{et al.} has emerged as the paradigmatic example of an isotropic potential that robustly forms icosahedral quasicrystals \cite{Engel2015}. Including three minima in the interaction potential is sufficient to observe the formation of the IQC over a broad region of parameter space, obtaining IQCs of varying density (correlating with the first coordination shell occupancy). Besides the IQCs, many complex periodic crystals are also obtained, including cP8, BC8, hP2, cP4, hP10 and clathrates. The IQCs are stable over long annealing, except in the high-density region, where the IQC transforms into an approximant crystal (AC)
\cite{Engel2015}. However, thermodynamic stability in the low- and intermediate-density region remains an open question, as spontaneous formation during cooling is not a definitive proof of stability. Complex energy landscapes are notorious for kinetic traps where metastable phases dominate on simulation timescales, especially when transitions to the more stable states require significant structural reconfiguration.

Determining phase diagrams for quasicrystal-forming systems has been hindered by the methodological challenges of simulating quasicrystal--fluid equilibria. Most attempts to do so have been limited to 2D systems \cite{Kiselev_Engel_Trebin_2012,Reinhardt13b,Reinhardt16,Dijkstra2017,Fayen2024,VarelaRosales24,Ulugol26} and typically involve some kind of approximation (e.g.\ applying some estimate for the configurational entropy of the quasicrystal). Demonstrating the thermodynamic stability of a quasicrystal is potentially somewhat simpler if spontaneous transitions to the quasicrystal from all relevant competing crystals can be observed, as has been evidenced for systems of hard tetrahedra that form a dodecagonal quasicrystal \cite{Je21}.

The two biggest challenges to computing the phase diagrams of quasicrystal-forming systems are the lack of a good reference state for free-energy calculations that use some form of thermodynamic integration \cite{frenkel1984}, and the perturbing effects of using periodic boundary conditions. Here, we circumvent the first challenge by using the ``direct coexistence'' method to directly simulate the equilibrium between the fluid and the quasicrystal \cite{Chew_Reinhardt_23}. This approach does, however, involve quasicrystalline slabs that span the periodic boundary conditions in two of the three dimensions. However, by considering boxes of different sizes, we show that the potential effects of these boundary conditions are small compared to the differences in stability allowing us to obtain confident predictions. Thus, we are able to show that for the OPP system the observed IQCs are not thermodynamically stable. Despite the high configurational entropy of the quasicrystals, they are consistently outcompeted by periodic phases such as hP2, BC8, and cP4, due to the significantly lower energy of these crystals.

\section{Method}

\subsection{The oscillatory pair potential}

We focus on systems interacting via the three-well oscillatory pair potential (OPP) proposed by Engel \emph{et al.} for which icosahedral quasicrystals were observed in simulations for the first time \cite{Engel2015}. 
This model was derived from a family of oscillatory potentials that mimic the Friedel oscillations that appear in the effective potentials of complex intermetallic compounds \cite{Mihalkovic_henley_2012}.
The OPP is based on the potential
\begin{equation}
V(r)=
    \frac{1}{r^{15}}+\frac{1}{r^3} \cos(k(r-1.25)-\phi)
\end{equation}
but where the potential is truncated and shifted at the third maximum, so that both the energy and the force vanish at this cutoff distance ($r_\mathrm{c}$). I.e.\
\begin{equation}
V_\mathrm{OPP}(r)=
    \begin{cases}
    V(r)-V(r_\mathrm{c}), & \text{if } r < r_\mathrm{c} \\
    0,       & \text{if } r \ge r_\mathrm{c} \\
\end{cases}
\end{equation}

The main effect of increasing the phase shift $\phi$ (in the range considered in Ref.\ \onlinecite{Engel2015}) is to increase the depth of the first minimum, with a small concomitant effect on the position and height of the first maximum (Fig.\ S1). 
An increase in the wavenumber $k$ leads to more rapid oscillations in the potential, but with the first maximum being roughly fixed at 1.25. This shifts the position of the second and third minima and the second maximum to shorter distances with the features becoming more pronounced due to the weaker attenuating effects of the $1/r^3$ factor. The first minimum is also significantly deepened. 

It will also be useful to calculate the contribution to the average potential energy per particle from pairs of particles with separations up to a given $r$. This is achieved by integration of the radial distribution function:
\begin{equation}
    v_\text{av}(r) = 2\pi \rho  \int_{0}^r r'^2 g(r') V_\text{OPP}(r') dr'.
\end{equation}

\subsection{Simulations}
All simulations are performed with LAMMPS \cite{LAMMPS}. The time-step used to integrate the equations of motion was 0.001, in reduced units. Temperature (and pressure) were controlled using the Nos\'{e}-Hoover thermostat and barostat, both with a relaxation time of 100 timesteps. 

\subsection{Direct coexistence method}

In the direct coexistence method \cite{Chew_Reinhardt_23}, a solid slab that spans the periodic cell in two of the three dimensions is placed next to a fluid.
When performed in the canonical ensemble, provided that the chosen state is not too far away from coexistence, the system can evolve by melting or freezing the solid in order to reach equilibrium, at which point the normal pressure becomes equal in the solid and fluid phases, thus determining the coexistence pressure at the chosen temperature. Although the solid can relax in the direction normal to the interface by expanding or contracting the atomic planes, it cannot do the same in the tangential directions because the lengths of the simulation box remain fixed. This potential accumulation of stress in the solid can be avoided by preparing an interface in which the dimensions in the tangential directions are chosen to match the normal pressure \cite{Davidchack_Laird_1998}. 
Alternatively, the coexistence conditions can be accurately estimated by finding the density at which the normal pressure in the direct coexistence simulations coincides with that of the bulk solid.\cite{Smallenburg2024} 
For simplicity, we have followed the first approach. We first perform a simulation to get an estimate of the coexistence pressure. Using this first estimate, a new fluid--solid interface is built with box dimensions that give the required lateral pressure.

In order for this approach to be successful, the fluid and the (quasi)crystal slab need to come into equilibrium. At equilibrium the rate of desorption from the slab will be equal to the rate of adsorption from the fluid. Thus, it is important that the simulations are long enough that a sufficient number of such adsorption and desorption events occur. Thus, as the coexistence pressure becomes small, longer simulations are required, simply because particle collisions with the surface are less frequent.

\begin{figure}
\includegraphics[width=45mm,angle=90]{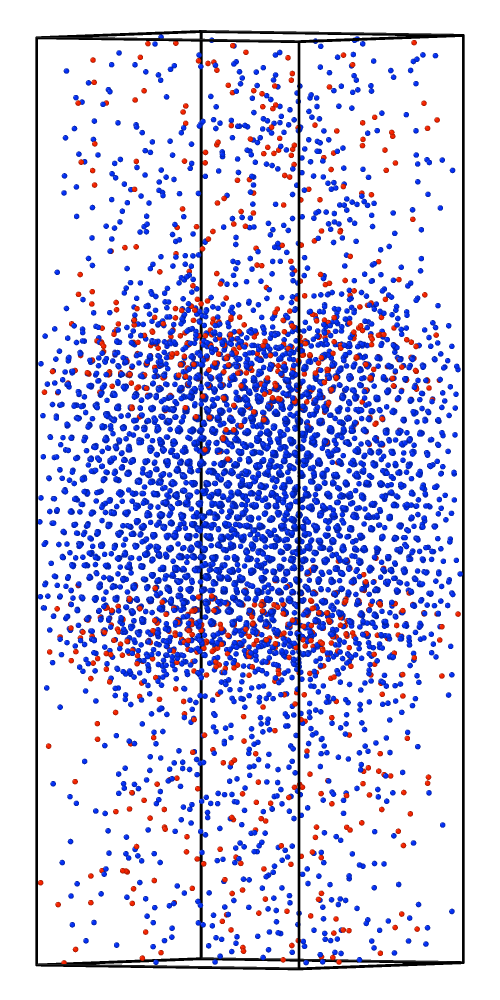}
\caption{\label{fig:I2-Interface} Fluid--IQC interface simulated with the I2 model at $T^*=0.3$. Particles that were in the fluid phase in the initial configuration are shown in red and those belonging to the IQC in blue. After 200 million simulation time steps, there has clearly been a significant exchange of particles between the two phases.}
\end{figure}

Furthermore, it is important that the surface of the slab reaches its equilibrium structure; this is particularly relevant when this is significantly different from the initial surface of the slab. In Fig.\ \ref{fig:I2-Interface} the exchange between a quasicrystal slab and the fluid is illustrated by colouring the particles by their initial identity as fluid or solid particles. It is noteworthy that the initially fluid particles are not just present at the IQC surface, but have infiltrated a significant depth into the slab, albeit not quite the very middle. This both reflects the relatively fast dynamics in this one-component isotropic-interaction system and the relatively high mobility of particles in a quasicrystal compared to a typical crystalline solid \cite{KaluginKatz,Bluher98}. 
It is also important that the solid slab remains sufficiently thick that there is a bulk-like region at the centre of the slab, so that any growth or melting of the slab leads to a change in thickness of that bulk-like region. This is particularly relevant for the long-ranged potentials that we consider here, as the long-ranged contributions are important for the energetic stability of the solids.

One might ask how the direct coexistence approach effectively samples the configurational entropy of a quasicrystal (the term that is problematic for thermodynamic integration (see Section \ref{sect:free_energy})). 
The somewhat trivial (and non-mechanistic) answer is that, if the fluid and solid have fully come into equilibrium, all contributions to the free-energy differences between the fluid and solid are effectively taken into account. However, considering the dynamical processes involved, the stabilizing effect of the configurational entropy must manifest itself as an increase in the rate of adsorption relative to desorption (compared to an equivalent system but without this entropic contribution). 
Although the dynamics at the surface are complex, one could envisage that the rate of adsorption is enhanced for a quasicrystal surface by the greater number of potential ways the particle could adsorb. 
Furthermore, often on surface adsorption, the adatom will not be at a particularly stable site, so may well desorb before it has been able to diffuse across the surface to locate a more stable location; the greater number of potential stable sites on the IQC surface may make such desorption less likely.
The situation is somewhat analogous to phase diagram calculations for ice $I_h$. The entropy associated with proton disorder needs to be explicitly added in free-energy calculations using thermodynamic integration, because simulations remain locked in a particular realization of the proton disorder, whereas direct coexistence simulations can directly determine the melting point\cite{Pauling1935,Sanz04,GarciaFernandez06,Piaggi20}. 

We note that direct coexistence simulations can also be performed at constant pressure \cite{Ladd1977}. In this approach, one runs simulations at a number of temperatures in order to identify the melting temperature at which the growth rate is zero. However, this approach can be quite slow, because as the melting temperature is more closely bracketed, one needs longer and longer simulations to ascertain whether the diffusive motion of the solid--fluid interface is leading to net growth or shrinking.

As in Ref.\ \onlinecite{Engel2015},
configurations for the IQCs were obtained from quenches of the fluid for each $(k,\phi)$ parameter set considered. 
Specifically, a low-density fluid ($\rho^*=0.03-0.05$) at a sufficiently high temperature (i.e.\ at a temperature at which the system is in the fluid phase) is gradually quenched in the canonical ensemble until spontaneous crystallization is observed. Under these conditions, a roughly spherical cluster in coexistence with a very low density fluid is always obtained. We then typically orient the cluster with three of its two-fold axes parallel to the simulation box and cut a cubic portion that is then allowed to relax in the $NpT$ ensemble. 

For the periodic crystals, an initial configuration was built simply by replicating the unit cell (typically with $N\approx$8000, which is sufficiently large to have negligible finite size effects) and equilibrating it at the desired pressure.

The solid block (periodic or aperiodic) is placed next to a fluid region about two or three times larger than the solid, where fluid particles are inserted randomly until a number density is achieved that is close to that of the fluid at the chosen conditions; note, a small gap is left between the fluid and the solid to avoid particle overlaps. A short simulation in the $NVT$ ensemble, keeping the solid fixed, is then performed to relax any overlaps in the fluid. 
Direct coexistence simulations were typically run for about 200--400 million time steps to ensure that the interface was fully equilibrated. 

It is important to consider the effects of the periodic boundary conditions on the IQC--fluid coexistence, as these break the aperiodicity of the quasicrystal.
Where the IQC meets itself across the periodic boundary conditions in the initial configuration, it is likely to lead to a grain-boundary-like feature (albeit with the same orientation either side of the boundary) where the local bonding is not as optimal as in the rest of the IQC grain. Even if the system is able to sufficiently anneal, so as to generate a uniform slab, it will still possess a finite phason strain associated with these two dimensions. (Phason strain is a measure of the deviation from perfect quasi-periodicity 
and in random tiling models of quasicrystals the tiling entropy is maximized as the phason strain goes to zero \cite{Henley91}.)  

\begin{figure}
\includegraphics[width=\linewidth,clip]{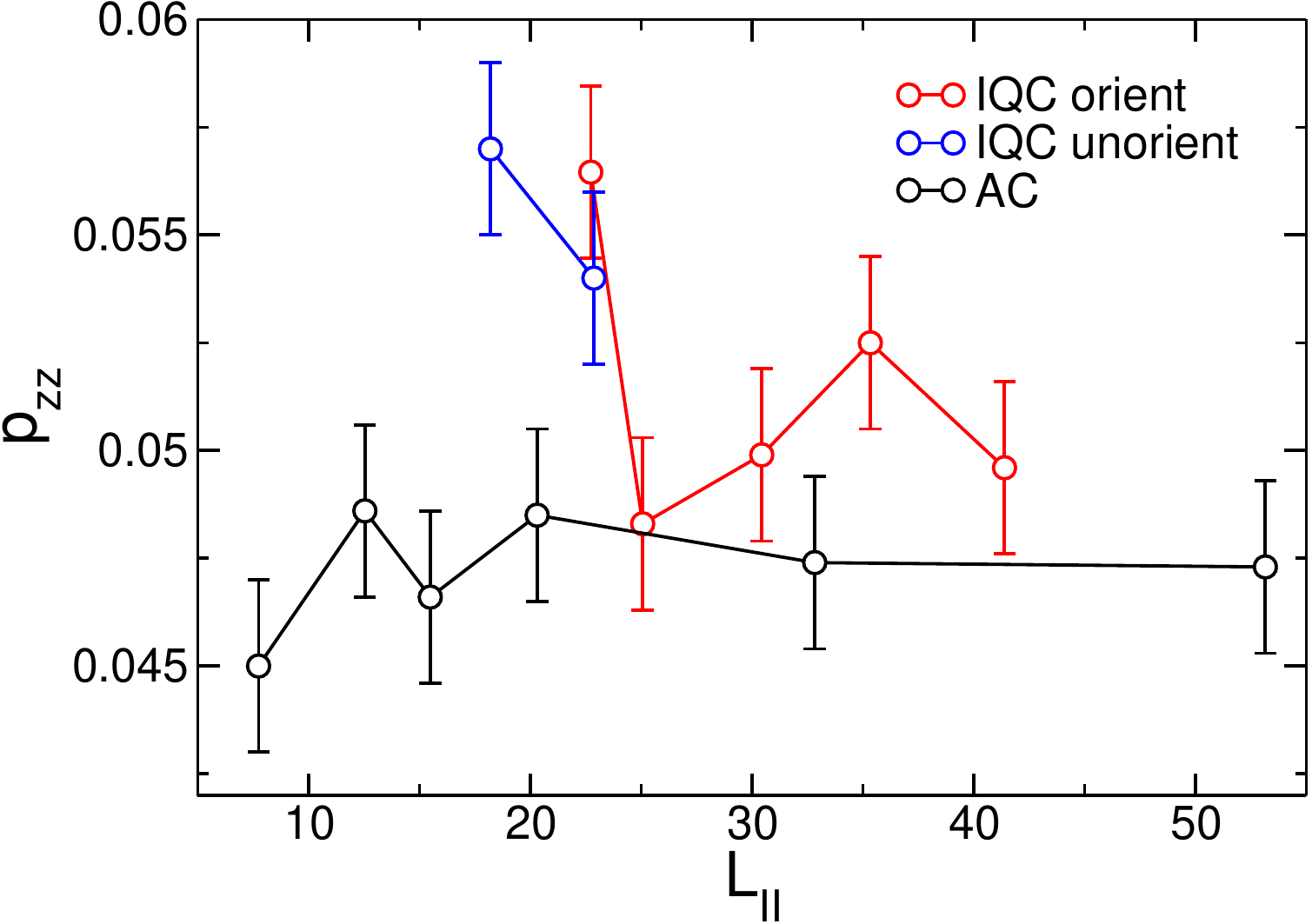}
\caption{\label{fig:IQC-fluid} IQC--fluid  and approximant--fluid coexistence pressure as a function of  the tangential box length, $L_{||}$, for the I1 model at $T^*=0.30$. }
\end{figure}

The adverse effects of the periodic boundary conditions on the IQC should reduce as the periodic dimensions increase, and so direct coexistence simulations were performed at multiple box sizes for the I1 model to assess these finite-size effects. The effects on the coexistence pressure are illustrated in Fig.\ \ref{fig:IQC-fluid}. The variation is of the order of 10--15\%. However, the variation is not uniform, because the statistical noise in the data is of a similar order.
We also considered the effect of the orientation of the IQC. The coexistence pressure for an IQC with its symmetry directions oriented randomly with respect to the box was not found to be significantly different. We also generated 
IQC approximants with increasingly large periodic repeats by projection from a 6-dimensional hypercubic lattice \cite{Noya21}. These also had similar coexistence pressures. Thus, for this example system (I1), we do not have sufficient accuracy to definitively comment on the relative stability of the IQC compared to its approximants. However, as we will see later, this accuracy is more than sufficient to draw definitive conclusions concerning the stability of the IQC with respect to other competing periodic crystals.

Of course, it would be preferable to be able to determine the IQC thermodynamics without the structure of the IQC being constrained by periodic boundary conditions. For example, in an $NVT$ simulation one can allow a large IQC cluster to come into equilibrium with a fluid. 
However, as the cluster grows or shrinks to reach this equilibrium, there is a contribution to the change in the free energy of the system that arises due to the change in the surface area of the IQC. Thus, separating bulk and surface contributions to the free energy (or the position of the equilibrium) is not straightforward. (N.b.\ The cluster in this equilibrium state actually corresponds to the critical nucleus \cite{MonterodeHijes20,MonterodeHijes22,Reguera26}.)
By contrast, in direct coexistence simulations, due to the slab geometry, the surface contribution to the total free energy remains constant as the IQC grows or shrinks. We note, though, that the bulk thermodynamics associated with defect formation in 2D quasiperiodic tilings has recently been extracted from simulations with open boundary conditions (an imposed line tension acts on the perimeter of the tile cluster) by performing a finite-size scaling analysis \cite{Ulugol25,Ulugol26}. Translating such approaches to particle-based systems, however, is not likely to be straightforward.

\subsection{Tracing coexistence curves with the Gibbs-Duhem method}
\label{sect:GibbsDuhem}

Once a coexistence point is known, it is possible to trace the entire coexistence line using the Gibbs-Duhem method\cite{Kofke93a,Kofke93b}, which consists of performing a numerical integration of the Clapeyron equation:
\begin{equation}
\frac{dp}{dT}=\frac{s_\text{sol}-s_\text{liq}}{v_\text{sol}-v_\text{liq}}=\frac{h_\text{sol}-h_\text{liq}}{T(v_\text{sol}-v_\text{liq})}
\label{eq:GDT}
\end{equation}
where $s_i$, $v_i$ and $h_i$ are the entropy, volume and enthalpy per particle of the two phases at coexistence. This first-order differential equation is numerically solved using a fourth-order Runge-Kutta method, with the enthalpy and volume of the two coexisting phases estimated from $NpT$ simulations.  In those cases in which the slope in the $p$--$T$ plane becomes too large, integration of the inverted equation is performed:
\begin{equation}
\frac{dT}{dp}=\frac{T(v_{sol}-v_{liq})}{h_{sol}-h_{liq}}
\label{eq:GDp}
\end{equation}
We typically use about 1 million MD steps to evaluate these properties at each stage of the Runge-Kutta integration. The increment in the independent variable ($T$ using Eq.~\ref{eq:GDT} and $p$ using Eq.~\ref{eq:GDp}) was adjusted depending on the slope of the coexistence line.

This method is especially useful for tracing the coexistence lines down to low temperatures for those phases whose coexistence pressure is extremely low (where equilibrating the vapour--solid interfaces becomes challenging). It is also used to extend the coexistence lines up to and beyond melting-point maxima; these occur when the volume of the coexisting liquid and solid are equal, making the slope of the coexistence line in the $p$--$T$ plane infinite. Such maxima are quite common for open low-density solids, such as those considered in this work, due to the higher compressibility of the fluid phase relative to the solid phase. 

\subsection{Calculation of the free energy and entropy}
\label{sect:free_energy}

 Simulations do not provide direct access to free energies. Calculation of the free energy requires the evaluation of the partition function; 
 this can only be achieved for a few reference systems for which the required multi-dimensional integration can be performed analytically.  Thus, free energies are often obtained using thermodynamic integration, i.e., by designing a continuous integration path between the system of interest and a reference system for which the free energy can be estimated analytically (or is known from previous calculations). For the fluid, a suitable reference system is the ideal gas, as it is usually possible to find a continuous integration path from any fluid to the ideal gas. For example, above the critical temperature, the free energy of the fluid, $A(\rho, T)$ can be estimated from\cite{frenkelbook}:

\begin{equation}
\frac{A(\rho, T)}{N k_{\text{B}} T} = \ln(\rho \Lambda^3) - 1 + \frac{\ln(2\pi N)}{2N} + \int_{0}^{\rho} \frac{Z(\rho')-1}{\rho'}  d\rho'
\label{eq:A_fluid}
\end{equation}
where $N$ is the number of particles, $\Lambda$ is the thermal de Broglie length, and $Z(\rho)=pV/Nk_\mathrm{B}T$ is the compressibility factor, which is a dimensionless quantity that measures how much a real gas deviates from the ideal gas.
In our calculations, $Z(\rho)-1$ is evaluated in $NpT$ runs and fitted to a polynomial function $f(\rho)=a_0\,\rho+a_1\,\rho^2+a_2\,\rho^3+a_3\,\rho^4+a_4\,\rho^5$ (the degree of the polynomial depends on the data set analyzed). This fit is inserted in the integrand of Eq.~\ref{eq:A_fluid} and the integral is resolved analytically. 

This approach combined with the results of the direct coexistence simulations allows the Helmholtz free energy of all solids (including the IQC) to be obtained. At coexistence the two competing phases have the same chemical potential, and so by performing thermodynamic integration from the ideal gas to the fluid at coexistence, the chemical potential of the coexisting solid is hence also obtained. Once the chemical potential is known, the Helmholtz free energy can be computed using:
\begin{equation}
A_\mathrm{DC}(p, T) = \mu_\mathrm{DC}(p, T) N -  p\,\langle V(p,T)\rangle,
\end{equation}
and also the total entropy:
\begin{equation}
 S_\mathrm{tot,DC}(p, T) = - \frac{A_\mathrm{DC}(p, T) - \langle U(p,T) \rangle}{T}.
 \label{Eq:S_tot}
\end{equation}

An alternative approach to obtain the free energy of a crystal is by thermodynamic integration from an appropriate reference state for the crystal. This has been performed (in two ways) both to provide a consistency check for the direct coexistence results and to explore whether the entropy of the IQC can be decomposed into configurational and vibrational components.

For crystals, 
the ideal Einstein crystal, which describes the crystal as a collection of non-interacting harmonic springs and whose free energy can also be computed analytically, is often used as the reference system. The solid is coupled to this reference system by gradually turning on the harmonic springs that bind each particle to its lattice site:
\begin{equation}
V(\lambda)= (1-\lambda) V_\mathrm{OPP} + \lambda (V_\mathrm{OPP}+ V_\mathrm{Eins})
\end{equation}
where $\lambda $ is a parameter that adopts values from 0 to 1.
To avoid a divergence in the integrand as the harmonic springs are switched off, translations of the solid need to be prevented. This can be achieved either by keeping the centre of mass fixed or by keeping the position of one particle fixed (see Refs.~\onlinecite{frenkel1984,Vega07,Vega08}). We opted for the latter, so that the free energy of the reference Einstein crystal is:
\begin{equation}
\frac{A_\mathrm{Eins}}{Nk_\mathrm{B}T}= \frac{1}{N} \ln\left(\rho\Lambda^3\right)+\frac{3}{2}\left(1-\frac{1}{N}\right)\ln{\left(\frac{ \Lambda_\mathrm{E}\Lambda^2}{\pi k_\mathrm{B}T}\right)}
\end{equation}
where $\Lambda_\mathrm{E}$ is twice the spring constant of the harmonic springs. 
The integration from the ideal Einstein crystal to the real crystal is performed in two steps. In the first step, the free energy between the non-interacting and interacting Einstein crystal is obtained by umbrella sampling:
\begin{eqnarray}
\frac{\Delta A_{1}}{Nk_\mathrm{B}T}  = &
-& \frac{1}{N} \ln\left\langle\exp \left[ - \left( V_\mathrm{OPP}-V_\mathrm{lat}\right)/k_\mathrm{B}T\right] \right\rangle_\mathrm{Eins} \nonumber \\
& + & \frac{V_\mathrm{lat}}{Nk_\mathrm{B}T}
\end{eqnarray}
where the angular brackets denote that the average is performed in a simulation of the Einstein crystal with one particle fixed, and $V_{lat}$ is the OPP energy of the crystal at its minimum.
In the second step, the free-energy difference between the interacting Einstein and real crystals is calculated by thermodynamic integration, along a path in which the harmonic springs are gradually turned off (again with one particle fixed):
\begin{equation}
\label{eq:A2}
\frac{\Delta A_{2}}{Nk_\mathrm{B}T}= -\frac{1}{N}\int_{\lambda=0}^{\lambda=1} \left\langle\frac{\Lambda_E\Lambda^2}{k_\mathrm{B}T} \sum_{i=2}^N(r_i-r_{i,0})^2\right\rangle_{\lambda} d\lambda.
\end{equation}

While this method has been extensively used for evaluating the free energy of periodic crystals, it cannot be directly applied to aperiodic lattices. 
In aperiodic crystals, besides the vibrational contribution to the entropy, there is a configurational term arising from all the different configurations that the quasicrystal can adopt which cannot be taken into account by the Einstein crystal method.

\begin{figure}
    \centering
    \includegraphics[width=0.8\linewidth]{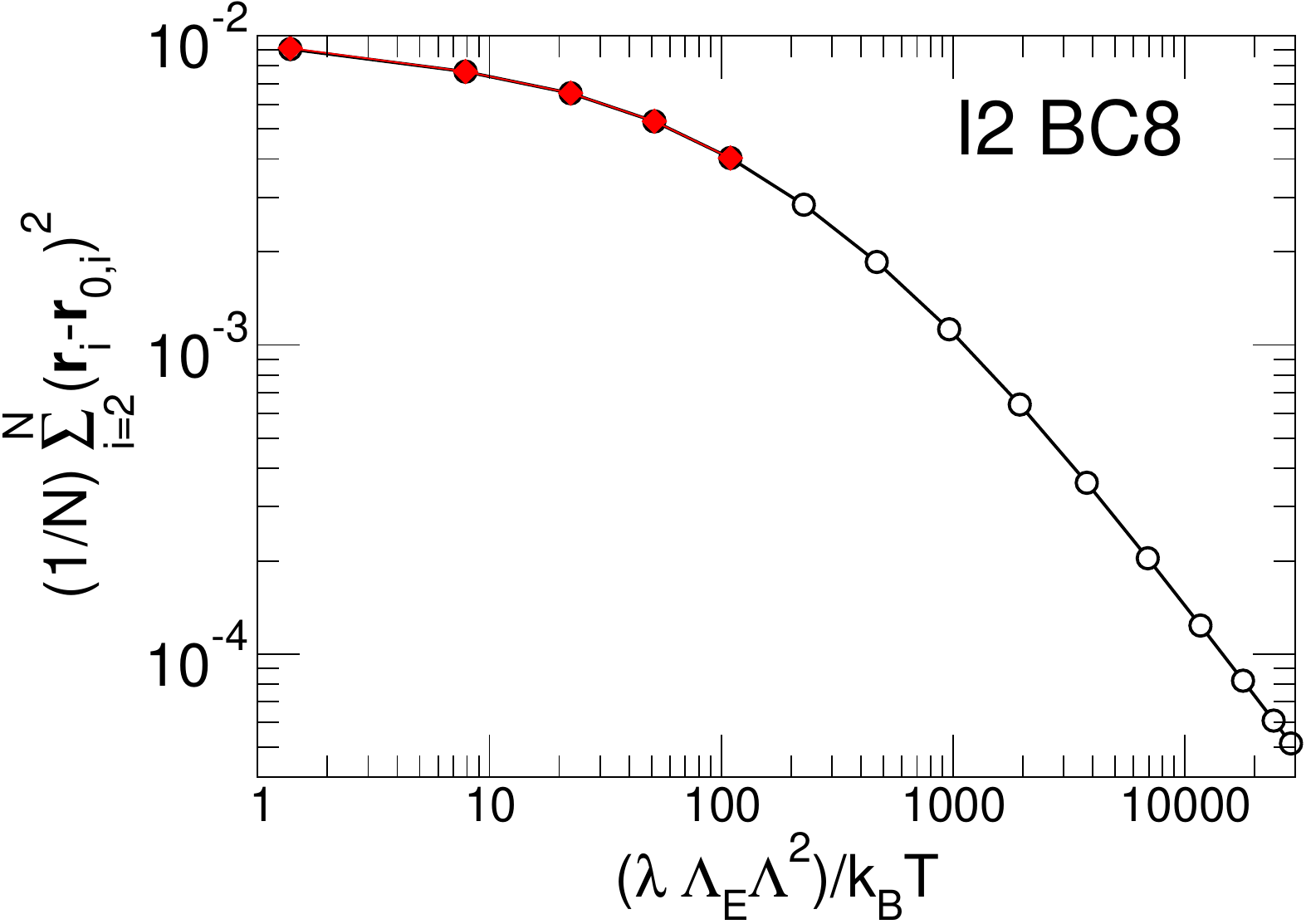}
    \includegraphics[width=0.8\linewidth]{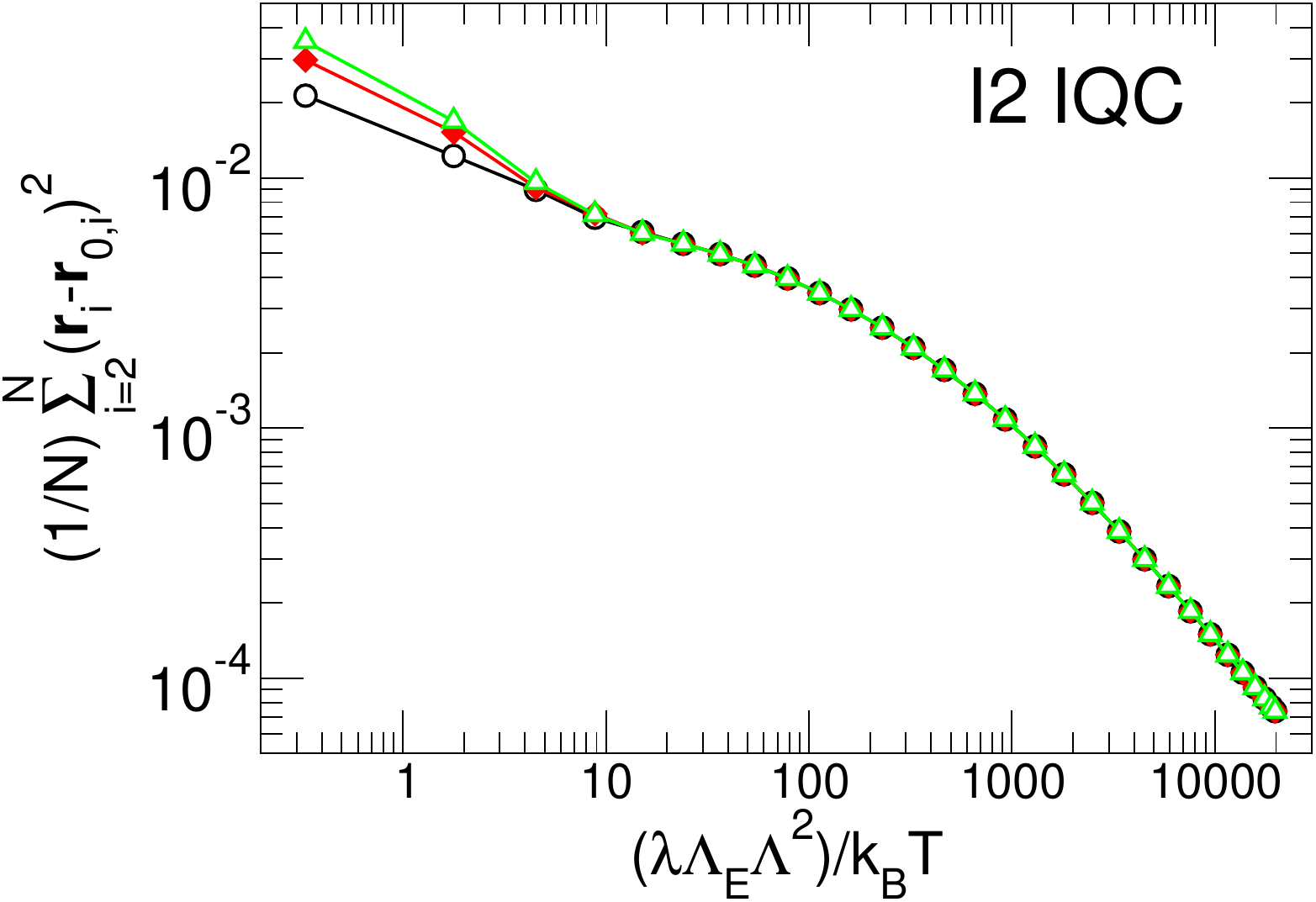}
    \caption{\label{fig:A2_integrand}Integrand of Eq.~\ref{eq:A2} as a function of $\lambda \Lambda_E\Lambda^2/k_\mathrm{B}T$ for the BC8 periodic crystal and the IQC. Simulations were performed for the I2 model at $T^*=0.27$ and coexistence densities, as estimated from the direct coexistence simulations. Black, red and green lines show the results of successive Monte Carlo runs consisting each of 400\,000 steps, with each run continuing from the final configuration of the preceding run. For low values of $\lambda$ the IQC is able to explore different energy minima (distinct configurations of the IQC), resulting in an increase in the integrand as $\lambda$ approaches zero. 
    }
\end{figure}

When applied to periodic crystals, the method relies on the simulations remaining in one permutational isomer of the crystal, even when the springs are turned off. If this condition also held for a quasicrystal it could be used to calculate the vibrational free energy of a quasicrystal configuration (as has been done a number of times \cite{Fayen2024,Dijkstra2017,Ulugol26}; note in Ref.\ \onlinecite{Ulugol26} additional constraints were applied to restrict the system to the starting configuration). However, for the systems studied here, other quasicrystalline configurations become kinetically accessible as $\lambda$ becomes relatively small and the mean-square displacement term in the integrand of Eq.\ \ref{eq:A2} begins to increase because of the onset of this diffusion. This effect is apparent in Fig.\ \ref{fig:A2_integrand}(b) for the IQC for $\lambda \Lambda_\mathrm{E} \Lambda^2/k_\mathrm{B}T$ less than about 10. In this range, the extent of diffusion, and hence the integrand's value, also depends on the length of the simulation. One might wonder whether Eq.\ \ref{eq:A2} could be used to calculate the full free energy of a quasicrystal if ergodicity could be maintained in the simulations for all values of $\lambda$ \cite{VarelaRosales24}. However, aside from numerical difficulties (e.g.\ the integrand will likely vary steeply as $\lambda\rightarrow 0$), this is problematic as explorations of permutational isomers of configurations, as well as different configurations, can contribute to the integrand.

Using a bespoke Monte Carlo code, we have calculated the free energies of the periodic crystals using this Einstein molecule approach ($A_\mathrm{EM}$) at a selection of the coexistence points identified by the direct coexistence simulations (see Table\ \ref{tab:A}). These are in very good agreement (with an average deviation of $0.01\,N k_\mathrm{B}T$) 
with the free energies computed by the first approach (i.e.\ $A_\mathrm{DC})$.

We also explore the use of the ``Debye crystal'' as an alternative reference system for the crystal. \cite{Koyama04,Manolopoulos11,ChengCeriotti2018, Chew_Reinhardt_23} The Debye crystal is a harmonic crystal whose vibrational frequencies and normal modes match those of the real crystal; the latter can be obtained by diagonalizing the Hessian matrix at the relevant potential energy local minimum.
The Helmholtz free energy of the Debye crystal is 
given by 
\begin{equation}
    A_\mathrm{h}(T,\rho) = V_\mathrm{min}(\rho)  + k_\mathrm{B} T \sum_{i=1}^{3N-3} \ln\frac{\hbar \omega_i}{k_\mathrm{B}T} 
\end{equation}
where $V_\mathrm{min}$ is the potential energy of the crystal at its local minimum and $\omega_i$ are the $(3N-3)$ vibrational frequencies. Note the Hessian has three zero-frequency eigenvectors that correspond to the translation of the centre of mass; if the centre of mass is not constrained a correction to the free energy must be included: $A_\text{com}/Nk_\text{B}T=-(1/N)\ln\rho$\cite{Polson_2000,Vega07,Vega08}; however, its contribution is negligible for the typical system sizes considered in this work.

\begin{figure}
    \centering
    \includegraphics[width=0.8\linewidth]{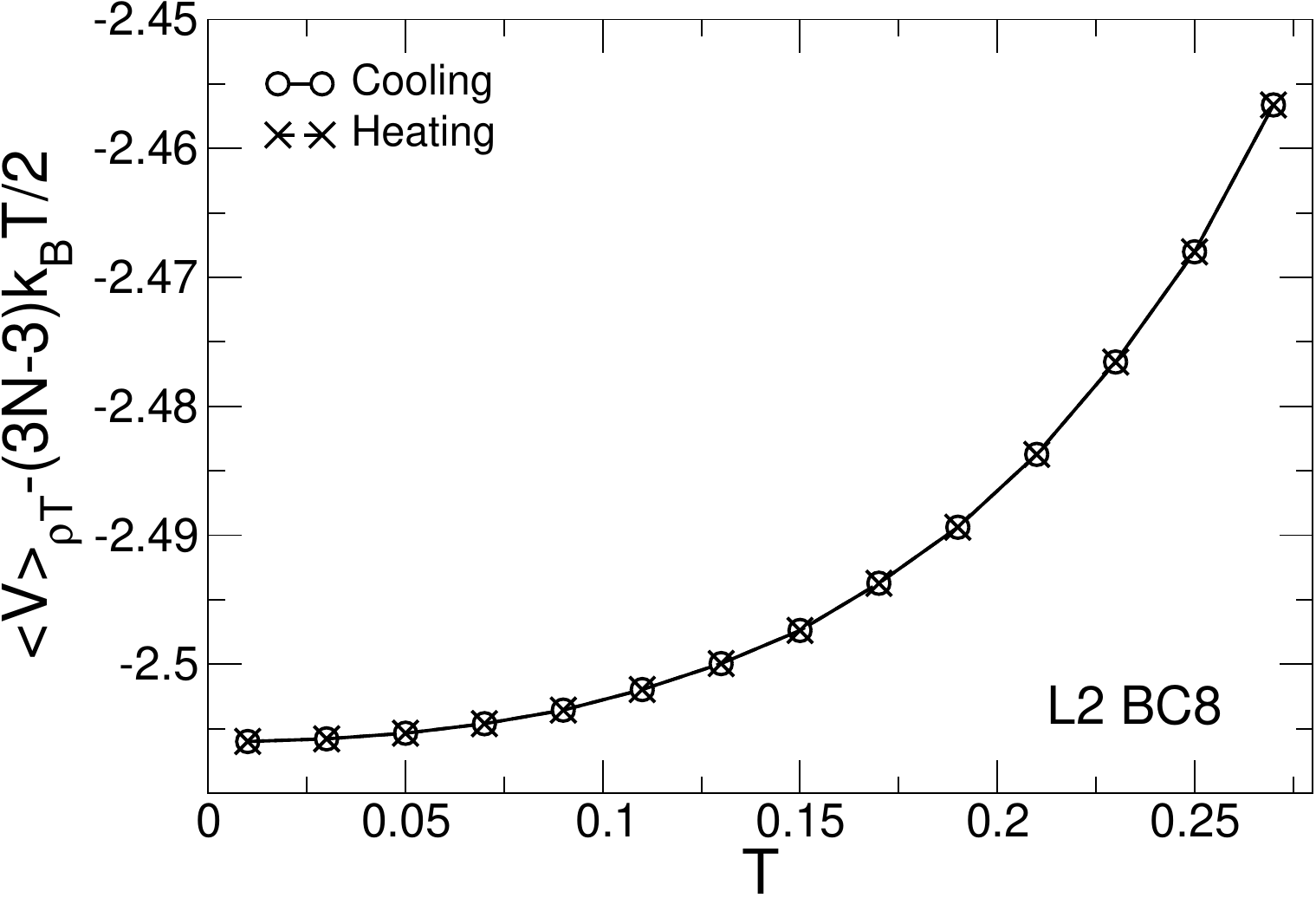}
    \includegraphics[width=0.8\linewidth]{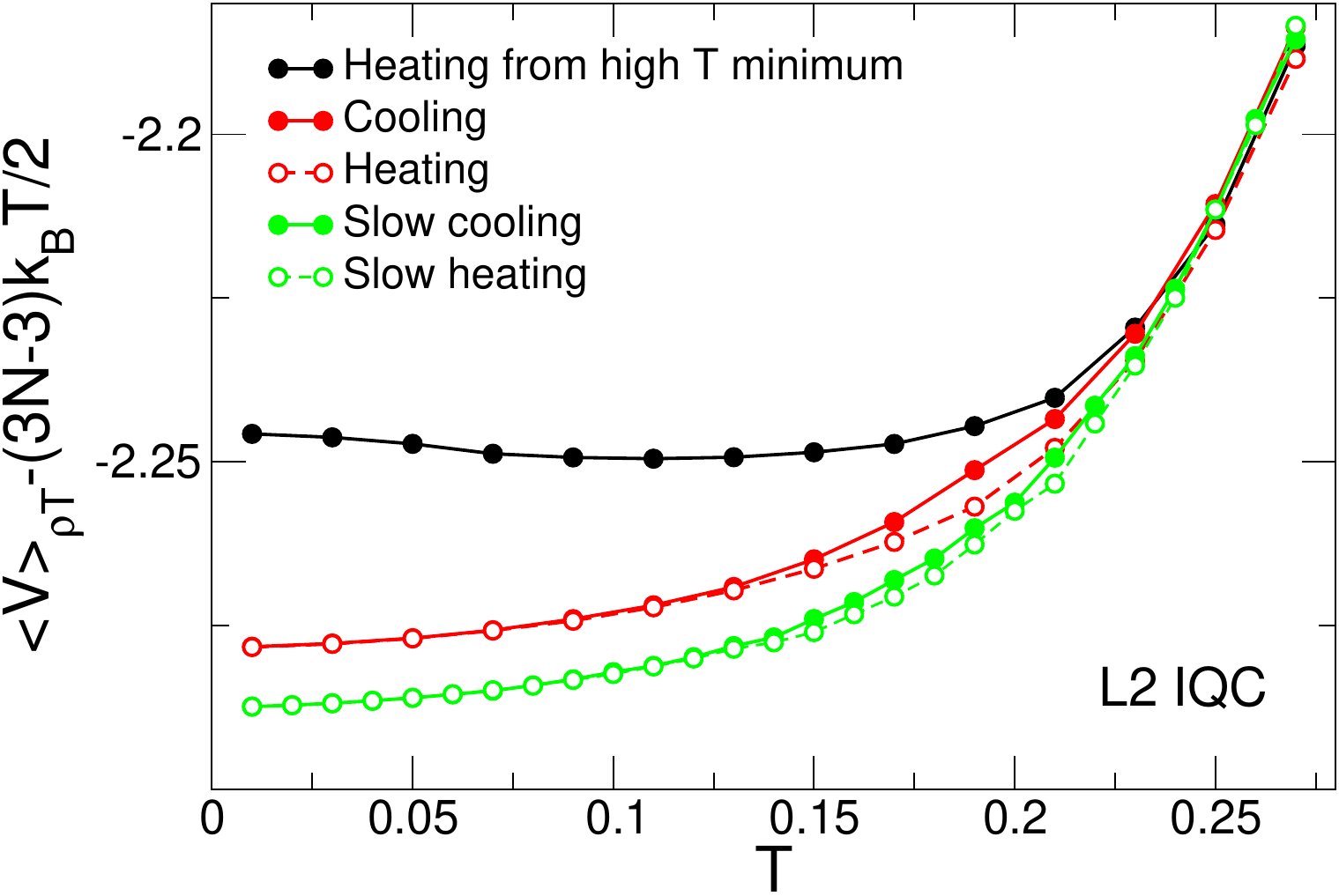}
    \caption{\label{fig:Debye_integrand} The potential energy (excluding the harmonic thermal contribution) as a function of temperature for 
    the BC8 periodic crystal (top) and  
    the IQC (bottom) in the L2 model at a density corresponding to that for coexistence at $T^*$=0.27. 
    The black heating line starts from a local potential energy minimum obtained from a simulation at $T^*$=0.27. Cooling and heating cycles with end points $T_1^*=0.01$ and $T_2^*=0.27$ were performed for the IQC at two different rates (red: 1 million MD steps per temperature; green: 10 million MD steps). 
    The slower quench is able to access lower energy minima. The quench-heating cycles exhibit some hysteresis, but this diminishes for longer simulation times. 
    }
\end{figure}

To compute the free energy of the real crystal, like in the Einstein crystal approach, we could perform a thermodynamic integration in which the Debye crystal Hamiltonian is transformed into that of the real system. However, at sufficiently low temperature, anharmonic effects are so small that the Helmholtz free energy of the Debye crystal is a very good approximation to that of the real crystal. Thus, this step can be skipped with little error. Instead, we now have to perform a thermodynamic integration to compute the free-energy difference of the crystal between this low temperature $T_1$ where the system effectively behaves harmonically and the temperature of interest $T_2$. The relevant integral is
\begin{equation}
\frac{A(\rho ,T_2)}{Nk_\mathrm{B}T_2} = \frac{A(\rho,T_1)}{Nk_\mathrm{B}T_1} -
\int_{T_1}^{T_2} \frac{\langle V+(3N-3)k_\mathrm{B}T/2 \rangle_{\rho,T}}{Nk_\mathrm{B}T^2} dT
\label{eq:inte_T}
\end{equation}
The integrand of this equation is often rewritten \cite{ChengCeriotti2018} in terms of the average anharmonic potential energy given by:
\begin{equation}
\langle V_\mathrm{a} \rangle_{\rho,T} = \langle V\rangle_{\rho,T} - V_\mathrm{min}(\rho) - (3N-3)k_\mathrm{B}T/2
\end{equation}
The last two terms correspond to the potential energy in the harmonic approximation. After some rearrangement, 
Eq.~\ref{eq:inte_T} can be written as:
\begin{eqnarray}
\frac{A_\mathrm{D}(\rho ,T_2)}{Nk_\mathrm{B}T_2} &=& \frac{V_\mathrm{min}(\rho)}{Nk_\mathrm{B}T_2} 
+ \frac{A(\rho,T_1)-V_\mathrm{min}(\rho)}{Nk_\mathrm{B}T_1} \nonumber \\
&& - \frac{(3N-3)}{N} \ln{\frac{T_2}{T_1}}  - 
\int_{T_1}^{T_2} \frac{\langle V_\mathrm{a}\rangle_{\rho,T}}{k_\mathrm{B} T^2} d T \nonumber \\
&=& \frac{V_\mathrm{min}(\rho)}{Nk_\mathrm{B}T_2} 
+  \frac{1}{N} \sum_{i=1}^{3N-3} \ln\frac{\hbar \omega_i}{k_\mathrm{B}T_1} \nonumber \\
&& - \frac{(3N-3)}{N} \ln{\frac{T_2}{T_1}}  - 
\int_{T_1}^{T_2} \frac{\langle V_a\rangle_{\rho,T}}{k_\mathrm{B} T^2} d T
\end{eqnarray}
Again, if the system explores only one energy basin, this route could be used to compute the vibrational entropy associated with that basin:
\begin{equation}
A_\mathrm{D}  = A_\mathrm{h} + A_\mathrm{a} = \langle U\rangle  - T S_\mathrm{vib}
\label{Eq:A_D}
\end{equation}
However, for simulations starting from a single IQC minimum we found that as the temperature increases the system is able to increasingly explore many different minima. Moreover, as the initial minimum is representative of the IQC at the coexistence temperature, this leads the system to relax to lower-energy IQC configurations that are more representative of the simulation temperature (see the black line in Fig.\ \ref{fig:Debye_integrand}).
Hence, if we wanted to use this approach to compute $S_\mathrm{vib}$ for an IQC configuration, we would need to artificially constrain the system to be associated with a single minimum.

The Helmholtz free energies of the crystals were computed via this approach at the same selection of coexistence points as considered for the previous methods. We chose $T_1^*$ to be 0.01; at this temperature the difference in internal energy between the real and harmonic systems is negligible.
The resulting values for the crystal free energies are again in good agreement with those obtained via the first two methods with an average deviation of $0.026\,k_\mathrm{B}T$ (see Table\ \ref{tab:A}). 
Our hypothesis for the somewhat lower accuracy is that numerical integration of Eq.\ \ref{eq:inte_T} starting at low temperatures increases statistical uncertainty. This is done by first fitting $\langle V_\mathrm{a} \rangle_{\rho,T}$ to a fourth- or fifth-order polynomial, and then solving the integral of Eq.\ \ref{eq:inte_T} analytically. The statistical error arises because the variation of the anharmonic energy across a wide temperature range cannot be fitted very accurately to a polynomial function.  

It is also interesting to look at the magnitudes of the anharmonic contribution to the Helmholtz free energy $A_\mathrm{a}$. These can be both positive and negative (note we are considering systems at constant density) and the largest magnitude of $A_\mathrm{a}/N$ that we obtain is $0.15\,k_\mathrm{B}T$, but they are frequently much less 
(the average magnitude across all solid phases is $0.06\,k_\mathrm{B}T$). This is useful as for the IQC, $A_\mathrm{DC}-A_\mathrm{h}=A_\mathrm{a}-TS_\mathrm{conf}$, where $S_\mathrm{conf}$ is the entropy associated with the number of distinct IQC configurations accessible to the system. In random-tiling models of quasicrystals \cite{Henley91}, this term is associated with the number of distinct random tilings. For particle-based systems, distinct configurations are often taken to be distinct minima on the potential energy landscape of the system \cite{Sciortino99a}. Thus, $(A_\mathrm{h}-A_\mathrm{DC})/N k_\mathrm{B}T$ provides an estimate of $S_\mathrm{conf}/Nk_\mathrm{B}$ of the IQC at coexistence to within an uncertainty of about 0.1. 
Since the IQC is able to explore different minima, $A_h$ was calculated by averaging over 100 minima obtained from minimizing configurations sampled at the coexistence temperature.

In principle, one could potentially compute the free energy of the IQC using the Debye approach if one started from the lowest-energy IQC configuration and was able to perform the simulations required for the thermodynamic integration along a reversible, equilibrium path. To explore to what extent this might be feasible, we slowly cooled down the IQC to $T_1$, and used the local potential energy minimum associated with the final configuration to compute $A_\mathrm{h}$. We then heated this configuration back up to $T_2$. From Fig.\ \ref{fig:Debye_integrand} we can see that there is a small amount of hysteresis between heating and cooling. This is because at some point during the cooling the system is no longer able to maintain equilibrium and fully relax to the lower-energy configurations relevant as the temperature is lowered. During the heating cycle, further relaxation can occur, slightly lowering the energy at intermediate temperatures compared to the cooling.

As the length of the cooling/heating runs increases the amount of hysteresis shrinks. More significantly, though, the longer runs are able to maintain equilibrium for longer during cooling, and so relax to significantly lower-energy regions of the IQC configuration space. Clearly, the estimates of the IQC free energy obtained from the simulations with different heating rates will be different, with the longer simulations leading to a lower estimate of the free energy because they are able to capture more of the IQC configurational entropy. However, the values obtained from the red and green heating curves in Fig.\ \ref{fig:Debye_integrand} are $0.21$ and $0.18\, N k_\text{B} T$  higher than that obtained by direct coexistence. Although these simulations can provide an upper bound to the quasicrystal free energy they are still a significant distance away from convergence.

The Debye approach effectively assumes that the starting state has zero configurational entropy (i.e.\ the single minimum dominates the equilibrium at $T_1$). However, the minimum obtained from the final configuration of the quench is instead representative of a set of minima with similar energies and configurational entropy $S_\mathrm{conf}(T_\mathrm{g})$, where $T_g$ is the effective temperature at which the system falls out of equilibrium. This is the source of the missing entropy for which the approach is unable to account. This situation is somewhat similar to supercooled liquids where the behaviour of the entropy below $T_g$ is generally inaccessible, leaving questions like the Kauzmann paradox \cite{Kauzmann48} hard to resolve.

Finally, we note that a final alternative way to compute the quasicrystal free energy might be to employ the Monte Carlo method developed by Schilling and Schmid for disordered or partially disordered systems\cite{SchillingSchmid09}, as it could potentially account for both the vibrational and configurational entropy. 
This method has been applied to a 2D quasicrystal in Ref.\ \onlinecite{Dijkstra2017}, although they mention that it seems to underestimate the configurational entropy for some of their quasicrystal-forming systems. Further exploration of this approach would be interesting, but is beyond the scope of the current paper.

We note that, for convenience, the values of the Helmholtz free energy that we report actually correspond to $A-Nk_\mathrm{B}T\ln\Lambda^3$,
obviating the need to specify the unit system in more detail. Consistent with this, 
the internal energies, entropies, enthalpies and chemical potentials that are reported are actually $U-3Nk_\mathrm{B}T/2$, $S-Nk_\mathrm{B}\left(3/2 - \ln\Lambda^3\right)$, 
$H-3Nk_\mathrm{B}T/2$ and $\mu-k_\mathrm{B}T\ln\Lambda^3$.

\section{Stability of the IQC}

\begin{figure}
\includegraphics[width=\linewidth,clip]{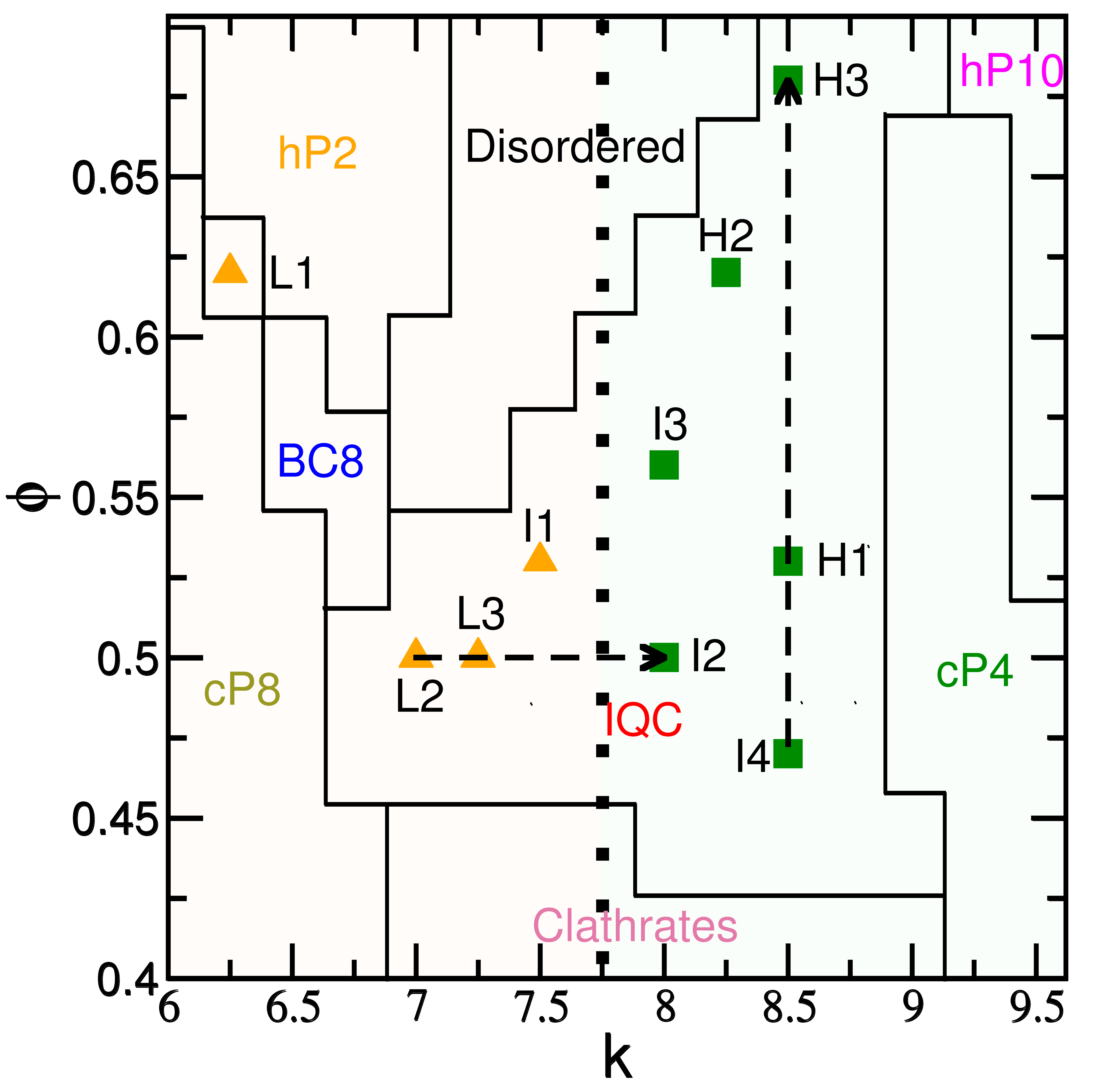} 
\caption{\label{Fig:PhaseDiagram_a} Self-assembly diagram dividing up the OPP parameter space by the solids that assemble on cooling a low-density fluid (replotted from Ref.~\onlinecite{Engel2015}). The $(k,\phi)$ parameter sets for which the fluid--solid coexistence was computed are marked by symbols and labels, following the same convention as in Ref.~\onlinecite{Engel2015}. The dashed arrows correspond to the paths considered in Sections \ref{sect:effect_k} and \ref{sect:effect_phi}. The vertical dotted line and background shading roughly divide up the same space by the identity of the thermodynamically most stable solid (hP2 to the left; cP4 to the right). }
\end{figure}

\begin{figure*}
\includegraphics[width=\linewidth,clip]{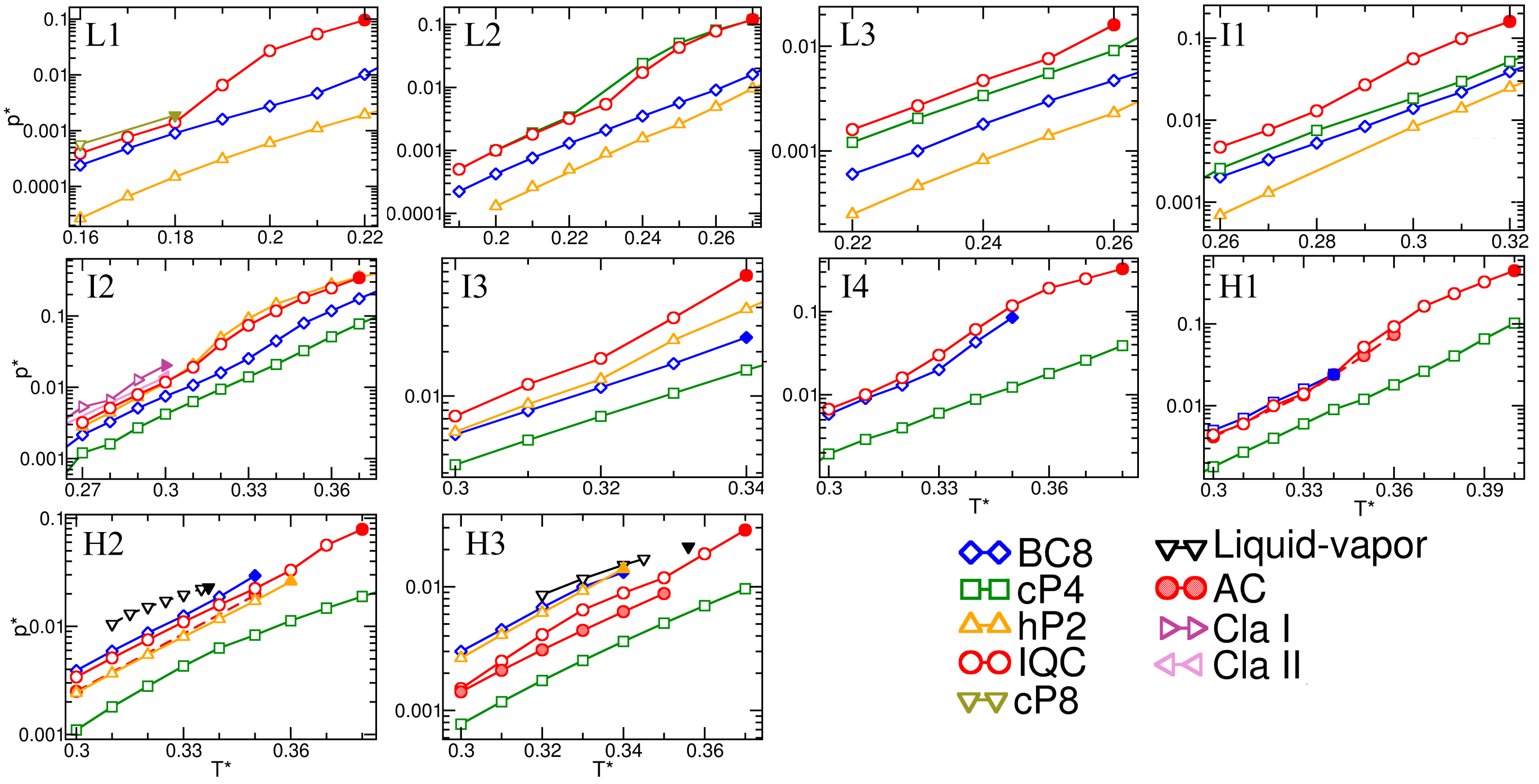} 
\caption{\label{Fig:PhaseDiagram_b} Pressure-temperature phase diagrams for the 10 models considered.  For each model, fluid--solid coexistence lines are included for those solid phases that were considered to be potentially competitive.
For the IQC, when these lines do not extend to the highest temperature it was due to the difficulty of stabilizing a solid--fluid interface at these temperatures.
For the liquid--vapour lines, the highest temperature point is the estimate of the critical point.
}
\end{figure*}

Fig.\ \ref{Fig:PhaseDiagram_a} shows the OPP parameter space considered in Ref.\ \onlinecite{Engel2015} divided into regions by the structures observed to assemble.
Direct coexistence simulations for the IQC and the competing periodic crystals were performed for 10 $(k,\phi)$ points in the low-density (L1, L2, L3), intermediate-density (I1, I2, I3, I4) and high-density (H1, H2, H3) regions previously identified by Engel \emph{et al.} to lead to IQC assembly \cite{Engel2015}. We note that investigations into the stability of the IQC for the L2 system have been reported in two PhD theses \cite{Je23,VarelaRosales24b}. These showed spontaneous transitions to the IQC for large clusters initiated from two quasicrystalline approximants \cite{Je23}, with approximate free-energy calculations showing that these transitions were driven by the greater entropy of the IQC despite its higher energy \cite{VarelaRosales24b}. Furthermore, simulations initiated with two clusters of different structure suggested that the IQC was unstable with respect to the BC8 and cI64 crystals \cite{Je23}.

Our results, summarized in Figs.\ \ref{Fig:PhaseDiagram_a} and \ref{Fig:PhaseDiagram_b}, reveal that despite the spontaneous formation of the IQC, a periodic crystal is thermodynamically more stable than the IQC for every parameter set considered (i.e.\ there is always a crystal that has a lower coexistence pressure with the fluid than the IQC). The differences between the coexistence lines of the IQC and the most stable periodic crystal are always sufficiently large to rule out the thermodynamic stability of the IQC, even assuming a 10--15\% overestimation of the coexistence pressure. As will be discussed later, the reason for the thermodynamic instability of the IQC stems simply from its relatively high energy. The energy difference between it and the most stable crystal is too large to be offset by any entropic advantage of the IQC.

\begin{figure}
\includegraphics[width=\linewidth,clip]{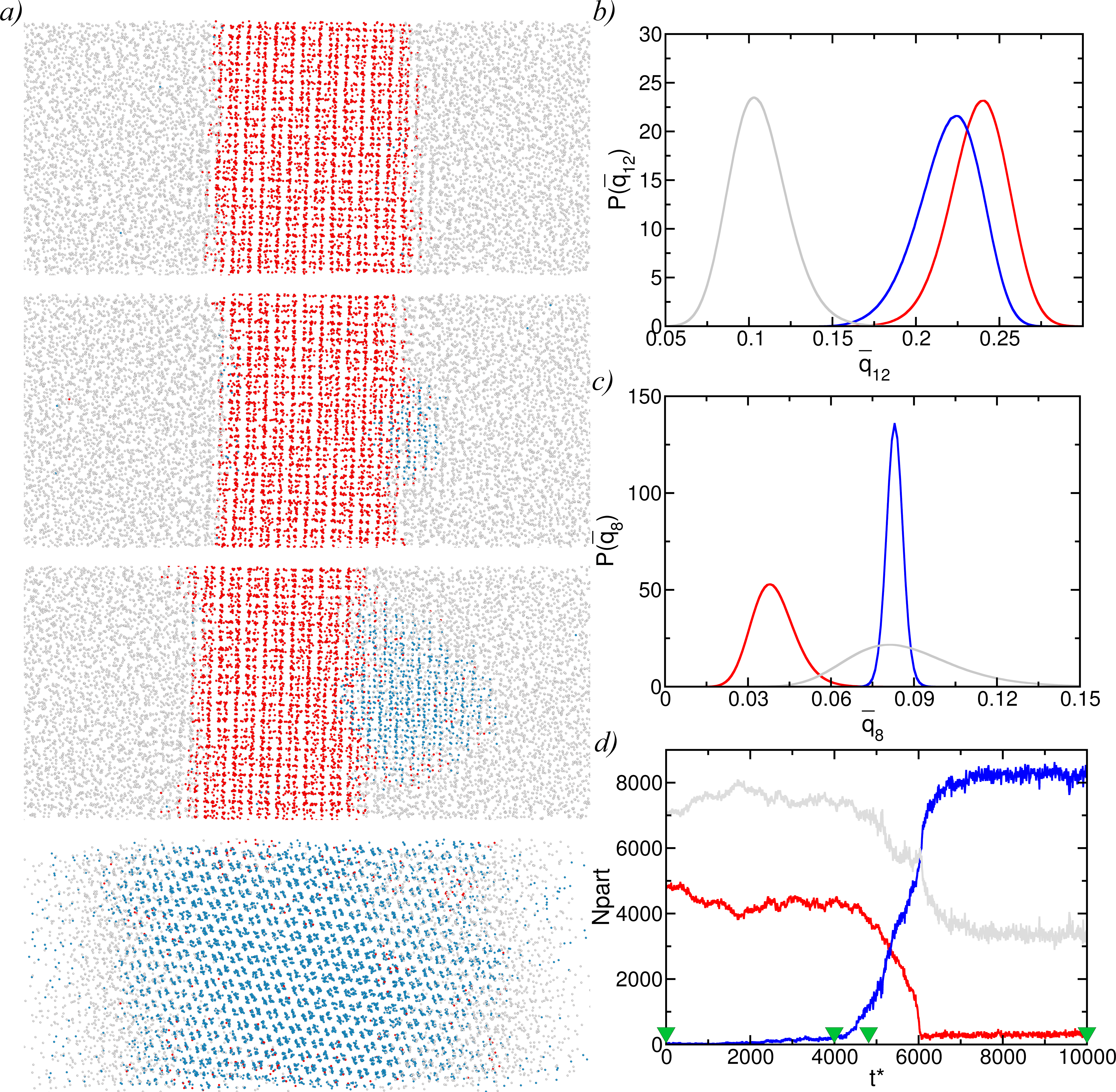} 
\caption{\label{Fig:L2-IQC-to-BC8}  Nucleation and growth of BC8 at the fluid--IQC interface for the L2 model at $T^*$=0.27. (a) Snapshots at different times show the equilibrated fluid--IQC interface ($t^*=0$), the nucleation of BC8 at one of the interfaces ($t^*=4\,700$) and its growth ($t^*=4\,750$) at the expense of both the fluid and IQC phases, and the final state ($t^*=10\,000$) in which the fluid coexists with BC8. (b) Particles in fluid or solid local environments can be distinguished using the local averaged $\bar{q}_{12}$ parameter\cite{Dellago2008}, with a threshold of $\bar{q}_{12} = 0.16$ being used. (c) Solid particles can subsequently be classified as BC8 or IQC using $\bar{q}_8$, with a threshold of $\bar{q}_{8} = 0.07$ being chosen. Both  $\bar{q}_{8}$ and $\bar{q}_{12}$ were calculated using a cutoff distance of 2.01 to define nearest neighbours. (d) Evolution of the number of particles in fluid, IQC and BC8 environments as a function of time. Green triangles indicate the time at which snapshots in a) are taken.}
\end{figure}

The instability of the IQC sometimes becomes evident in the direct coexistence simulations. For example, for one of the L2 model simulations, the fluid--IQC interface undergoes a spontaneous transition to fluid--BC8 after long simulation times (Fig.~\ref{Fig:L2-IQC-to-BC8}). 
This fluid--IQC interface was equilibrated at $T^*=0.27$ and $p^*=0.12$. Under these conditions, the IQC coexists with a fairly dense fluid ($\rho_\mathrm{f}^*\approx0.365$). Analysis of the trajectory reveals that the IQC-BC8 transition starts by the heterogeneous nucleation of a BC8 seed at the fluid--IQC interface, which grows both at the expense of the fluid and IQC phases, eventually converting into a fluid--BC8 interface. 
The observed structural transformation of the IQC contrasts somewhat with the stability of L2 IQC clusters over very long tempering simulations reported by Engel \emph{et al.}\cite{Engel2015}, however these simulations were at lower temperature and much lower pressure.

In the high-density region, the 1/1 approximant identified by Engel \emph{et al.}\cite{Engel2015} (Fig.\ S3) appears to be more stable than the IQC for the H3 model, but it loses some stability for the H1 model. In the latter case, the IQC and AC exhibit comparable stabilities within the uncertainty of our simulations. 

Interestingly, among all the solid phases considered, only the cP4 and hP2 crystals exhibit thermodynamic stability for the $(k,\phi)$ points considered. This is somewhat surprising, since the regions where these crystals were observed to assemble spontaneously from the fluid are quite distant from some of these points. The wavenumber $k$ seems to be the key parameter; we find that hP2 is the stable phase for $k\lesssim7.25$, whereas for $k\gtrsim8.0$ cP4 becomes the most stable. Curiously, some of the phases that spontaneously formed showed relatively little stability for the considered parameter sets. For example, cP8 is the least stable solid for L1. 
Similarly, clathrates I and II are the least stable phases among those considered for I2. The high energy of these phases explains (at least in part) their low  thermodynamic stability.

We have also calculated the liquid--vapour coexistence lines for the H2 and H3 models.
This was achieved by performing direct coexistence simulations (starting from a high-density fluid slab in contact with a vacuum) at several temperatures and using the law of rectilinear diameters to estimate the critical point\cite{frenkelbook}. We found that the liquid--vapour critical points are in the region of the phase diagram in which the cP4 crystal is most stable. Furthermore, for H2, the liquid--vapour coexistence line lies above all the fluid--solid coexistence lines in the $p$--$T$ phase diagram. 

The relative instability of the liquid phase is a consequence of the positive regions in the OPP pair potential. Pair distances corresponding to these regions are much more prevalent in the liquid than the observed crystals leading to the relative energetic destabilization of the liquid. We did not compute the liquid--vapour coexistence lines for any of the L or I models because the liquid is likely to be even less stable for these models due to the increasingly positive nature of these potentials (particularly for pair distances corresponding to the first-neighbour shell).

\begin{figure}
\includegraphics[width=\linewidth,clip]{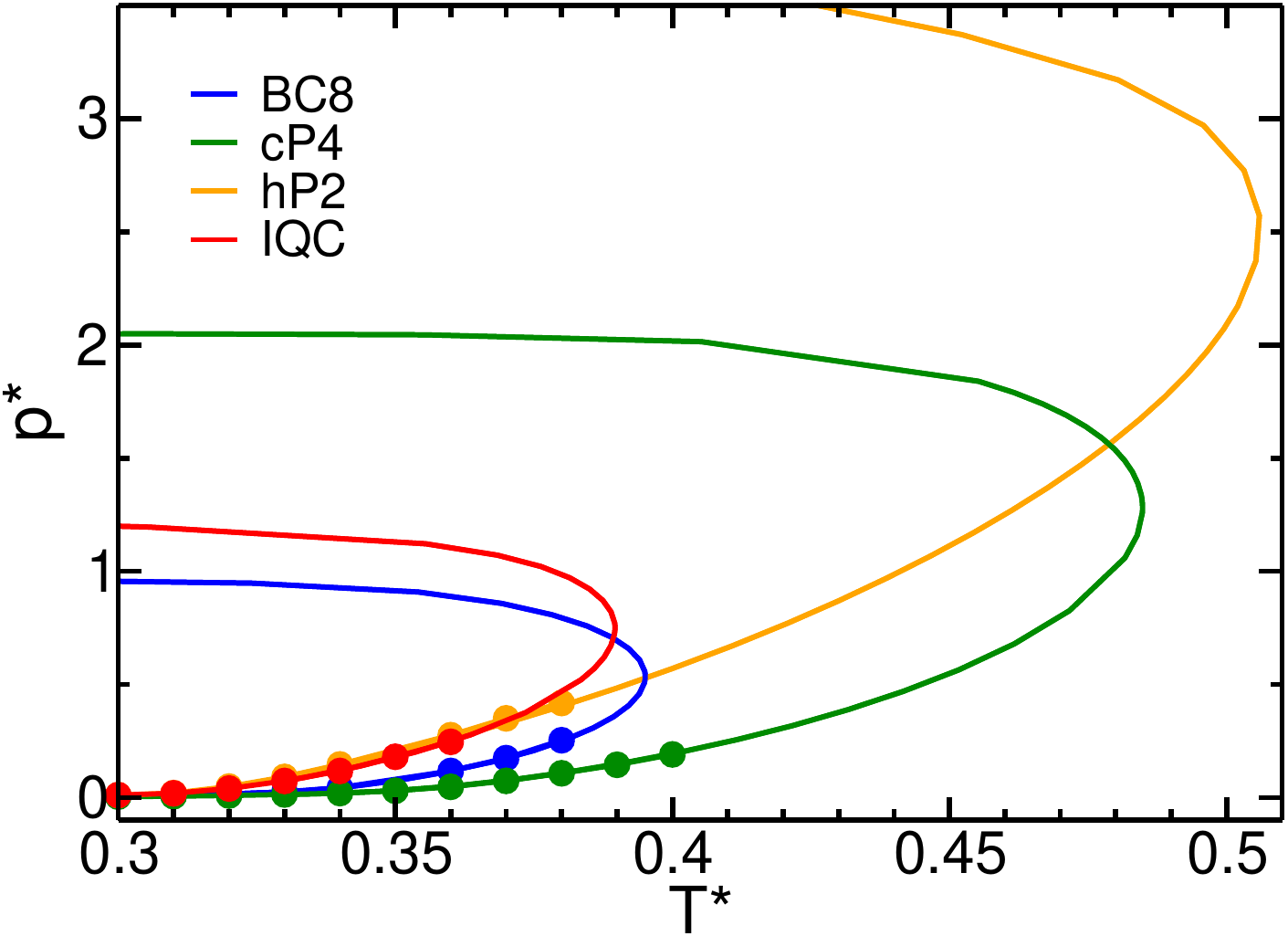} 
\caption{\label{Fig:reentrant} Pressure-temperature phase diagram for the I2 model. Fluid--solid coexistence lines were extended beyond the temperature range considered in Fig.\ \ref{Fig:PhaseDiagram_b}
using Gibbs-Duhem integration. All coexistence lines exhibit a melting-point maximum (where the fluid and the solid exhibit the same density). Filled circles were obtained with the direct coexistence method, whereas lines correspond to results obtained with Gibbs-Duhem integration.}
\end{figure}

In Fig.\ \ref{Fig:PhaseDiagram_b}, temperatures were considered up to the point where we were no longer able to form a stable IQC--fluid interface in the direct coexistence simulations. It is not that the phase boundary ends at these points, but that the system is approaching a melting-point maximum where $dT_m/dp=0$ and the phase boundaries reverse direction. Such melting-point extrema occur at the point where the difference in density between the coexisting IQC and fluid is zero. At temperatures less than that for the maximum, the IQC is only stable with respect to the fluid for a finite range of density, and there is a second phase transition line at higher pressure corresponding to the transition from the IQC to a higher-density fluid (see Fig.\ \ref{Fig:reentrant}). 
Melting-point maxima are not uncommon for energetically stabilized low-density crystals and have been seen in patchy-particle systems \cite{Noya07b,Romano12}. for repulsive potentials with multiple length scales \cite{Prestipino05,Fomin08} and for multiple-feature potentials that, like $V_\mathrm{OPP}$, have both attractive and repulsive features \cite{Fomin11}.  They are also present in the experimental melting lines of graphite and the alkali metals.\cite{Togaya97,vandeWalle19}

For the I2 model,  we have used the Gibbs-Duhem approach to trace the coexistence curves of all the competing solids to pressures higher than that for the melting-point maxima (see Fig.~\ref{Fig:reentrant}). All the solids exhibit a melting-point maximum (BC8 and IQC in a metastable region with respect to cP4). In principle, locating the higher-pressure sections of these lines could have also been done by direct coexistence, but where the fluid in contact with the crystal/IQC has a higher density. The presence of a triple point in this phase diagram where cP4, hP2 and the fluid coexist also suggests there is a transition from cP4 to the higher-density hP2 at higher pressures, but we have not explored whether there might be any other competing solids phases in this region of state space.

An interesting observation is that for the IQC the melting point maximum occurs at somewhat higher temperatures than those for which we could stabilize the IQC--fluid interface. There are several potential reasons why it might become more difficult to equilibrate the fluid--solid interfaces as the melting-point maximum is approached. First, as the densities of the fluid and solid become more similar, the range of system densities at which it is possible to stabilize the interface becomes narrower. Second, it is possible that for a finite system a homogeneous phase is most stable, even when the overall density corresponds to the two-phase region, because the free-energy cost of an interface is not insignificant.\cite{Binder12} For example, if the density is just a bit higher than that for the fluid binodal, the stable state of a finite system can still be a homogeneous fluid. 
Third, one needs to be more careful that the solid slab does not become destabilized by becoming too thin.

%
%

\subsection{\label{sect:effect_k} Effect of the potential parameter $k$}

\begin{figure*}
\includegraphics[width=\linewidth,clip]{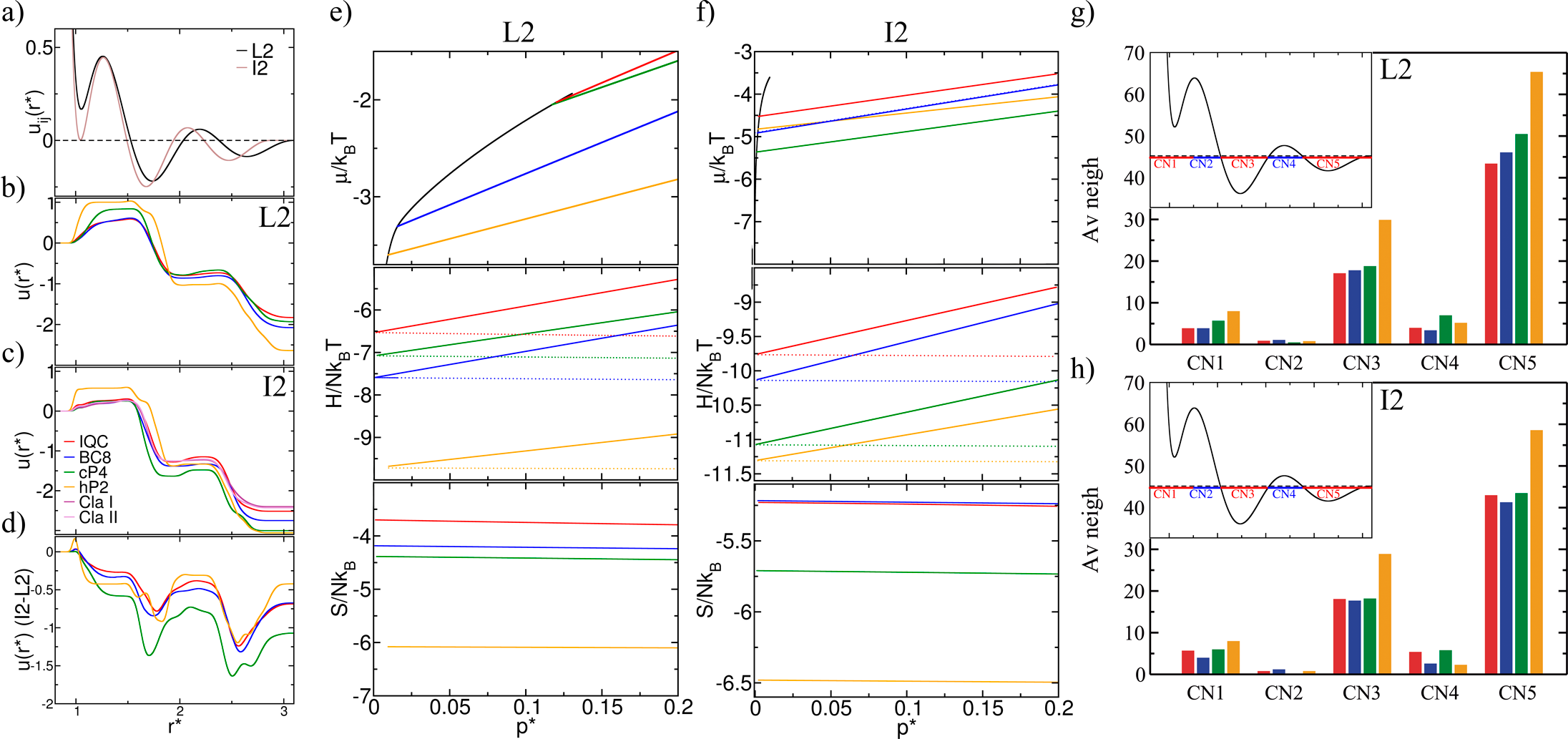} 
\caption{\label{Fig:thermo-L2-I2} Effect of increasing the wavenumber $k$ at constant $\phi$ in the model potential: from L2 to I2. (a) Functional form of the L2 and I2 models. (b) and (c) Integrated potential as a function of distance for several solids for L2 and I2, respectively. The difference between the integrated potentials is shown in (d). (e)and (f) Chemical potential of the fluid and the competing solids (top panel), as well as the enthalpy (middle panel) and entropy (bottom panel) for (e) L2 and (f) I2 at $T^*=0.27$. (g) and (h) Average coordination number in the regions as defined in the insets for (g) L2 and (h) I2 The integrated potentials and average coordination numbers were calculated at $T^*=0.27$ and the corresponding coexistence pressure. 
}
\end{figure*} 

To analyse the thermodynamics of these systems in more detail, we first consider the three systems L2, L3 and I2, which correspond to increasing $k$ whilst keeping $\phi$ constant at 0.5. Increasing the wavenumber, $k$, in the potential function leads to faster oscillations in the interactions. The first minimum becomes significantly deeper remaining roughly at the same distance, whereas the second and third minima shift toward shorter distances and become slightly deeper. The resulting potential models for L2 and I2 are shown in Fig.\ \ref{Fig:thermo-L2-I2}(a).

For the L2 model ($k=$7.0), hP2 emerges as the most stable solid, followed by BC8, whereas the IQC and cP4 phases are notably less stable, exhibiting nearly degenerate melting points. This stability hierarchy largely correlates with the average energies  (see Tables \ref{tab:A} and S2): hP2 possesses the lowest energy, while BC8 ($u_\textrm{BC8}\approx 0.78\,u_\textrm{hP2}$) remains significantly more stable than cP4 ($u_\textrm{cP4}\approx 0.73\,u_\textrm{hP2}$), and the IQC is the least energetically favourable phase ($u_\textrm{IQC}\approx0.68\,u_\textrm{hP2}$).

Further insights can be obtained from the integrated potential energy per particle as a function of distance (Fig.\ \ref{Fig:thermo-L2-I2}(b)). Interestingly, the energy of hP2 remains higher than that of the other solids up to the first coordination shell. This is easily understood by the positive value of the first potential minimum, combined with the significantly larger coordination number of hP2 (8.0 neighbours) compared to the alternative phases (3.5--5.0) (Fig.~\ref{Fig:thermo-L2-I2}(g) and Table S3). However, once the second and third coordination shells are taken into account, hP2 exhibits the lowest energy configuration. This occurs because the coordination numbers for distances corresponding to the second and third attractive wells in the potential significantly exceed those of the competing solids. 

The hP2 phase is also favoured by its efficient packing, with a number density of approximately 0.9, compared to the 0.55--0.68 range of the competitive solids (Table \ref{tab:A} and S2). However, given the low melting pressures, the $pV$ contribution to the Gibbs free energy plays a minor role in the relative stabilities at conditions close to melting. 

Interestingly, despite the significantly lower energy of hP2, its melting pressure is only slightly lower than that of BC8, which suggests that the entropy of BC8 is significantly higher than that of hP2. 
This is confirmed by our free-energy calculations (derived from the direct coexistence results and by equating the chemical potential of the fluid and the solid, see Section\ \ref{sect:free_energy}), which together with the measured enthalpies, allow the entropy to be obtained. 
As shown in Fig.\ \ref{Fig:thermo-L2-I2}(e), BC8 indeed exhibits a significantly higher entropy than hP2. It also surpasses that for cP4, leaving the IQC as the only solid with an entropy higher than that of BC8. Similarly, the IQC displays a stability similar to cP4. Despite its higher energy, it is stabilized by its higher entropy relative to the cP4 phase.

As the wavenumber increases towards the I2 model ($k=$8.0), all the solid phases gain stability with respect to the fluid phase (i.e., the melting curves shift to lower pressures). This is a consequence of all the potential minima becoming deeper, especially the first one. The energy gain is significantly larger in the solid than in the low-density fluid at coexistence, which explains the enhanced stability of the solids as $k$ increases. 
Significantly, the relative stability of the competing solids is reshuffled. cP4 is now the most stable solid, followed by hP2 and BC8 which exhibit similar chemical potentials, while the IQC remains as the least stable phase.

\begin{table*}
\centering
\small
\caption{\label{tab:A}Thermodynamic properties of the solid phases at coexistence with the fluid for a selection of systems and temperatures. These include free energies obtained with thermodynamic integration from the Debye crystal, direct coexistence simulations and the Einstein Molecule method. 
All free energies $A_{i}$ are given in units of $Nk_\mathrm{B}T$ and the entropy in units of $Nk_\mathrm{B}$; note, 
$U_\mathrm{min}$ and $\langle U\rangle$ are given per particle in reduced units.
$S_\mathrm{tot}/Nk_\mathrm{B}=(\langle U \rangle-A_\mathrm{DC})/Nk_\mathrm{B}T$. 
Some of the entries for the IQCs have been left blank as the quantities are not well-defined but vary with simulation length due to partial exploration of the IQC configurational space, as discussed in Sect.\ \ref{sect:free_energy}. 
}
\begin{tabular}{cccccccccccccccc}
\toprule
Model & Phase  & $T^*$ & $\rho^*$ & $p_{zz}$ & $\langle U\rangle$ & $U_\mathrm{min}^*$ & $A_\mathrm{h}$ & $A_\mathrm{a}$ & $A_\mathrm{D}$ & $A_\mathrm{DC}$ & $A_\mathrm{EM}$ & $A_\mathrm{DC}-A_\mathrm{EM}$ & $A_\mathrm{DC}-A_\mathrm{D}$ & $A_\mathrm{h}-A_\mathrm{DC}$ & $S_\mathrm{tot}$ \\
\midrule
L2    & BC8    & 0.27  & 0.5724   & 0.0160   & $-2.051$ & $-2.506$      & $-3.286$  & $-0.106$ & $-3.392$  & $-3.407$   & -3.396   & $-0.011$          & $-0.013$  & 0.121  & $-4.189$  \\ 
L2    & hP2    & 0.27  & 0.9020   & 0.0096   & $-2.624$ & $-3.028$      & $-3.619$  & 0.023 & $-3.596$  & $-3.640$   & $-3.634$   & $-0.006$          & $-0.044$  & 0.021  & $-6.079$  \\
L2    & cP4    & 0.27  & 0.6761   & 0.118    & $-1.920$ & $-2.321$      & $-2.723$  & 0.058 & $-2.665$  & $-2.687$   & $-2.684$   & $-0.003$          & $-0.022$  & $-0.036$  & $-4.424$ \\
L2   & IQC    & 0.27  & 0.5522   & 0.120    & $-1.780$ &  $-2.239$ & $-2.277$ & & & $-2.829$   &          &                 & & 0.552    & $-3.764$ \\
\midrule
I2    & BC8    & 0.27  & 0.6482   & 0.0021   & $-2.737$ & $-3.181$      & $-4.819$  & $-0.091$ & $-4.910$  & $-4.922$   & $-4.909$   & $-0.013$ & $-0.012$  & 0.103   & $-5.215$ \\
I2    & hP2    & 0.27  & 0.9647   & 0.0028   & $-3.054$ & $-3.404$      & $-4.641$  &  $-0.034$ & $-4.675$ & $-4.651$   & $-4.622$   &  $-0.029$  & $-0.024$ &  0.010  & $-6.482$ \\
I2    & cP4    & 0.27  & 0.7635   & 0.0012   & $-2.991$ & $-3.403$      & $-5.399$ &  $-0.024$ & $-5.423$  & $-5.451$   & $-5.456$   & 0.005    & $-0.028$    & 0.052  & $-5.627$ \\
I2    & IQC    & 0.27  & 0.7276   & 0.0032   & $-2.639$ &  $-3.064$   & $-4.086$ & & & $-4.536$   &  &  & & 0.450  & $-5.238$ \\
\midrule
I4    & BC8    & 0.33  & 0.6765   & 0.020    & $-2.749$ & $-3.316$      & $-3.075$  & $-0.136$ & $-3.211$  & $-3.237$   & $-3.233$   & $-0.004$ & $-0.026$  & 0.162  & $-5.093$ \\
I4    & cP4    & 0.33  & 0.7943   & 0.006   & $-3.231$ & $-3.738$ & $-4.073$  & $-0.068$ & $-4.141$  & $-4.104$   & $-4.105$   & $-0.001$          & 0.037   & 0.031                & $-5.687$ \\
I4    & IQC    & 0.33  & 0.7181   & 0.030    & $-2.739$ & $-3.253$  & $-2.688$   & & & $-3.096$   &          &                 & & 0.408    & $-5.204$ \\
\midrule
H3    & BC8    & 0.33  & 0.6564   & 0.0100   & $-2.882$ & $-3.453$      & $-3.603$  & $-0.151$ & $-3.754$  & $-3.716$   & $-3.738$   & 0.022           & 0.038   & 0.113                & $-5.017$ \\
H3    & hP2    & 0.33  & 0.9615   & 0.0089   & $-3.391$ & $-3.894$      & $-3.727$  & $-0.036$ & -3.763  & $-3.795$   & $-3.771$   & $-0.024$          & $-0.032$   & 0.068   & $-6.481$ \\
H3    & cP4    & 0.33  & 0.7712   & 0.0026   & $-3.468$ & $-3.976$      & $-4.881$  & $-0.027$ & $-4.908$  & $-4.894$   & $-4.916$   & 0.022           & 0.014   & 0.013                & $-5.615$ \\
H3    & IQC    & 0.33  & 0.7692   & 0.0065   & $-3.109$ & $-3.629$  & $-3.718$   & & & $-4.063$   &          &                 & & 0.345    & $-5.358$ \\
\bottomrule
\end{tabular}
\end{table*}

The energetic advantage of hP2 diminishes as a consequence of the changes in the potential profile. 
While hP2 maintains its high first-shell coordination (8 nearest neighbours), the shift of the second and third minima to shorter distances excludes a significant proportion of potential neighbours from the interaction range (about 10\%). In the L2 model, these particles would have fallen within the third attractive region for hP2, but they now lie outside the interaction range. In contrast, cP4 adapts better to the changes in the potential function. In the I2 model, cP4 not only increases its first-shell coordination (from 5.7 to 6.0) but also effectively empties its first repulsive region, with the number of neighbours dropping from 0.5 to 0.1. The integrated potential energy difference between the L2 and I2 models (Fig.\ \ref{Fig:thermo-L2-I2}(d)) evidences that a significant fraction of the energy gain occurs at distances lower than 1.25, precisely where cP4 better accommodates its first coordination shell to avoid the first potential maximum.
Consequently, the energetic difference between hP2  and cP4 in the I2 model becomes almost negligible, with the energy of hP2 now being only marginally lower than that of cP4.
While BC8 and the IQC also reduce their energetic disadvantage relative to hP2, they do so to a lesser extent. The bulk of the energy gain in these phases arises mainly from pair interactions within the 2.7–3.10 range, again due to the aforementioned loss of interacting pairs in hP2 within the third minima. This finding underscores the relevance of the third well in the potential for the relative stability of the competing solids.

 Given that, as before, the $pV$ term does not play a significant role at the low coexistence pressures, it is the larger entropy of cP4 that renders it more stable than hP2 (Fig.~\ref{Fig:thermo-L2-I2}(h)). The same applies to BC8, which exhibits a higher energy than hP2 ($u_\mathrm{BC8}\approx0.92\,u_\mathrm{hP2}$), but due to its highly favourable entropy it is more stable than hP2. Perhaps surprisingly, for I2, BC8 and the IQC exhibit virtually the same entropy (for L2, the IQC was the solid phase with highest entropy). 
 Given that there is a configurational contribution to the entropy of the IQC (coming from all the possible realizations of the IQC structure) which is absent in BC8, these results show that for I2 the vibrational entropy of BC8 is sufficiently large to match the combined vibrational and configurational entropy of the IQC. Although the relationship between structure and vibrational entropy is going to be complex for a long-ranged multi-feature potential like the OPP potential, we note that the BC8 crystal has the lowest density of the solids considered at I2, and this may be part of the reason why the motion of the atoms is less constrained. 

 The trend towards increasing stabilization of cP4 with increasing $k$ is also seen for the intermediate model L3, but, as this point is much closer to L2 than I2, the hP2 crystal is still comfortably the most stable phase.

\subsection{\label{sect:effect_phi} Effect of the potential parameter $\phi$}

\begin{figure*}
\includegraphics[width=\linewidth,clip]{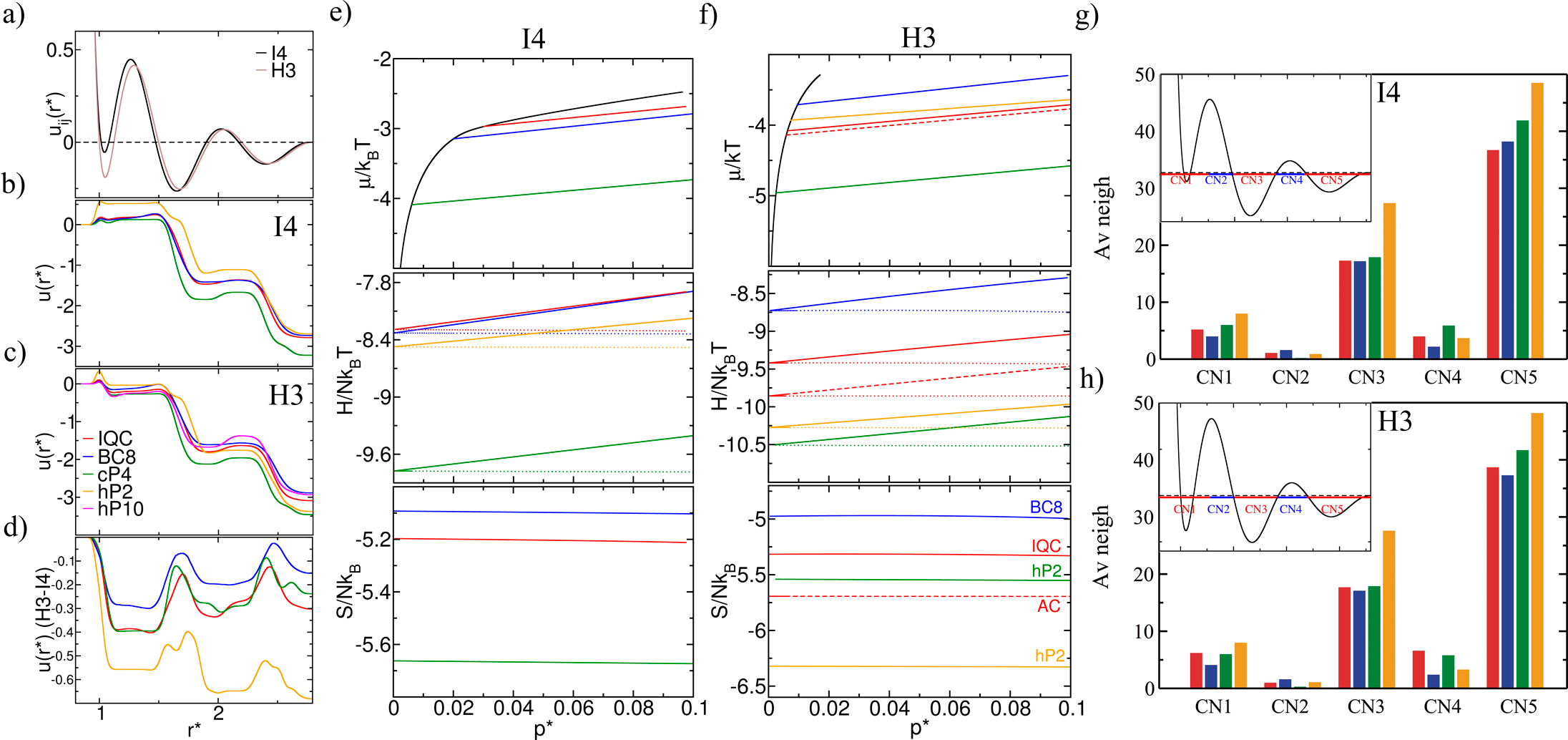} 
\caption{\label{Fig:thermo-I4-H3} Effect of increasing the phase $\phi$ at constant $k$ in the model potential: from I4 to H3. (a) Functional form of the I4 and H3 models. (b)-(c) Integrated potential as a function of distance for several solids for I4 and H3, respectively. The difference between the integrated potentials is shown in (d). (e)-(f) Chemical potential of the fluid and the competing solids (top panel), as well as the enthalpy (middle panel) and entropy (bottom panel) for (e) I4 and (f) H3 at $T^*=0.33$. (g)-(h) Average coordination number in the regions defined in the insets for (g) I4 and (h) H3. The integrated potentials and average coordination numbers were calculated at $T^*=0.33$ and the corresponding coexistence pressure. 
}
\end{figure*}

To consider the effect of the parameter $\phi$ on the solid--fluid thermodynamics, 
we next consider the solid--fluid thermodynamics of the three systems I4, H1 and H3 in more detail, as these correspond to increasing $\phi$ while keeping $k$ constant at 8.5.
When $\phi$ is increased, the main change in the potential is that the first minimum becomes deeper. Other subtler changes include a small reduction in the height of the first maximum, and a small shift toward longer distances for all the potential extrema. This is illustrated in Fig.\ \ref{Fig:thermo-I4-H3}(a) for the models I4 and H3. 

For the I4 model, the most stable phase is cP4, followed by BC8 and IQC, whereas we were unable to form a stable hP2--fluid interface at the temperatures considered. Interestingly, the ordering of the phases according to their energy is cP4 $<$ hP2 $<$ BC8 $\lesssim$ IQC. Although we were unable to stabilize the fluid--hP2 system at $T^*=0.33$, the bulk hP2 crystal remains mechanically stable and exhibits an average energy slightly lower than that of BC8 and IQC, both of which have very similar energies. 
The less favourable entropy of hP2 likely leads to the melting-temperature maximum for hP2--fluid coexistence occurring at relatively low temperature.
Another interesting observation is that BC8 is now the phase with the highest entropy, slightly surpassing the IQC and significantly exceeding cP4.

As $\phi$ increases, the energy of all polymorphs is lowered (see Fig.~\ref{Fig:thermo-I4-H3}(d) and Tables \ref{tab:A} and S4). However, the magnitude of the energy change is again not the same for all of the solids. At $T^*$=0.33, the polymorph that benefits the most from the deepening of the first minimum is hP2 (which has a coordination number of 8, compared to 6 in cP4 and 4 in BC8), with its energy reduced by 20\%. Significant reductions are also observed for the IQC (13\%), followed by cP4 (7\%) and BC8 (7\%). A marked difference between the IQC and the periodic crystals is that its first coordination shell becomes more populated when the first minimum becomes deeper, going from about 5.2 neighbours for I4 to about 6.2 for H3 (see Fig.\ \ref{Fig:thermo-I4-H3} and Table S5). Interestingly, the larger energy gain of the IQC compared to cP4 is also due to an increase in the number of neighbours in the third minimum in the IQC. 
Even though packing does not play a significant role in stability under the conditions considered, it is interesting to note that the IQC and the periodic crystals behave differently. The IQC becomes denser as $\phi$ increases, whereas the periodic solids become slightly less dense (by about 2–3\%). 

As a consequence of these energy shifts, significant changes are observed in the phase diagrams. First, all the fluid--solid coexistence lines move to lower pressures due to the gain of energy of the solid phases when $\phi$ increases. Both for I4 and H3, the most stable phase is cP4 due to its significantly lower energy. hP2 does not remain stable in coexistence with the fluid phase for the I4 model (neither for the H1 model), transforming into the most stable cP4 after a short simulation time (Fig.\ S6). Only for H3 is hP2 stable in contact with the fluid for the temperature range considered. The reason is that hP2 has the largest number of neighbours in the first coordination shell (coincident with the first minima), which is barely negative for I4, but becomes more negative as $\phi$ increases. The situation is different from L2 or I2 because for those models, the higher energy of the first coordination shell is compensated by the favourable contributions from the second and third minima. 

The relative stability of BC8 and IQC reverses with increasing $\phi$, due to the greater energy gain of the IQC. While BC8 is slightly more stable than the IQC for the I4 model, the IQC becomes the second most stable phase of those considered for H3 in Fig.\ \ref{Fig:PhaseDiagram_b}. Interestingly, for both sets of parameters, BC8 exhibits the largest entropy, slightly above that of the IQC. This again indicates that the vibrational entropy of BC8 must be larger than the combined vibrational and configurational entropy of the IQC. 

\begin{figure*}
    \centering
    \includegraphics[width=0.9\linewidth]{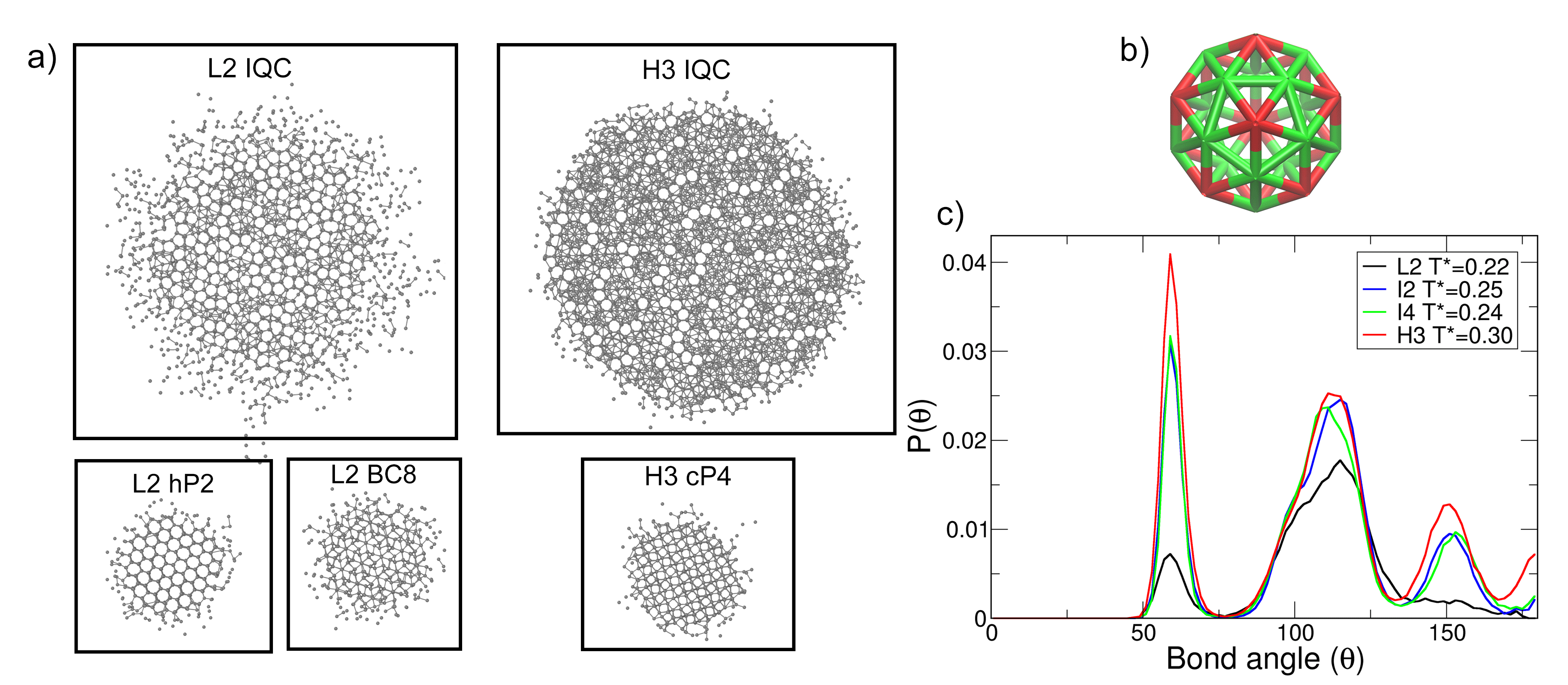}
    \caption{(a) Slices of thickness 5 from solid clusters formed in self-assembly simulations. Particles in the surrounding fluid have been removed and lines are drawn between particles that are less than 1.3 apart. The IQCs are viewed down one of the five-fold axes. (b) The rhombic triacontrahedral unit. The twelve red vertices are arranged in an icosahedron, while the twenty green vertices in a dodecahedron. (c) Angular distribution functions associated with nearest-neighbour bonds (within a cutoff distance of 1.3) for IQCs for the L2, I2, I4 and H3 systems, normalized to the average coordination number. Average coordination numbers are 3.45 for L2, 5.36 for I2, 5.24 for I4 and 6.28 for H3 (at the temperatures shown in the legend). 
    }
    \label{fig:IQC_structure}
\end{figure*}

\subsection{Structural evolution of the IQC}
One interesting feature of the IQCs is that their structure can evolve as a function of conditions to a much greater extent than typical for a crystal. For example, as identified in Ref.\ \onlinecite{Engel2015}, as $k$ and to a lesser extent $\phi$ increase, the density of the IQC increases significantly with  a consequent shift to higher coordination numbers. At higher densities, icosahedral clusters with a rhombic triacontahedral structure are an evident feature of the IQC (Fig.\ \ref{fig:IQC_structure}). These clusters have two environments, namely, those associated with a 12-particle icosahedron where the edge length matches that of the second minimum in the OPP potential and those associated with a 20-particle dodecahedron where the 
edge length is the nearest-neighbour distance. As the first minimum in the potential becomes less favourable, these dodecahedral sites become depopulated, reducing the density and the average coordination number. These changes are also reflected in the angular distribution function associated with nearest-neighbour bonds. Angles near 60$^\circ$ and 150$^\circ$ that are associated with this dodecahedral environment become less prevalent (Fig.\ \ref{fig:IQC_structure}(c))\cite{Noya21}. 
The directionality of the nearest-neighbour bonds in the IQCs has been previously represented using bond-orientational order diagrams (BOODs) \cite{Engel2015}. The bonds between the particles in the dodecahedral sites occur along the 2-fold directions of the Ih point group \cite{Noya21}, and these were also clearly less present in the BOODs of the low-density IQCs.

It is also interesting to examine how these changes affect the configurational entropy of the IQC. Although we were not able to fully isolate the configurational component of the entropy, as mentioned in Sect.\ \ref{sect:free_energy}, $(A_\mathrm{h}-A_\mathrm{DC})/Nk_\mathrm{B}T$ gives an estimate of $S_\mathrm{conf}/Nk_\mathrm{B}+A_\mathrm{a}/Nk_\mathrm{B}T$ at the coexistence temperatures. As $A_\mathrm{a}$ is not likely to vary much between the IQCs at different density and the values of $A_\mathrm{a}$ for the crystals suggest $A_\mathrm{a}/Nk_\mathrm{B}T$ is not likely to contribute much more than 0.1, one can deduce from the values of $(A_\mathrm{h}-A_\mathrm{DC})/Nk_\mathrm{B}T$ that $S_\mathrm{conf}$ decreases significantly as the density of the IQC increases (Table \ref{tab:A}). This trend is consistent with the structural features noted above. At high density, well-ordered rhombic triacontahedral clusters are present, whereas at lower density there is significant disorder associated with the occupation of the dodecahedral sites. This illustrates that configurational entropy in quasicrystals not only stems from different possible ``tilings'' (idealized IQCs can be made from two types of cell called Ammann rhombohedra \cite{Senechal04,Kramer84}), but also from disorder in the decoration of these tiles.

An important question raised by the results presented so far is why structures (including the IQCs) that are significantly higher in free energy than the most stable solid phase readily form in self-assembly simulations. Although there are many examples where the initial nucleation of metastable phases has been identified \cite{Ostwald,Sanz07,Zubieta19,Noya19,Donadio25},
the current systems are quite extreme in this regard. For example, of the ten $(k,\phi)$ parameter sets considered only for H1 was the thermodynamically most stable phase observed and then only 30\% of the time \cite{Engel2015}. A common way to interpret this type of behaviour is through classical nucleation theory\cite{kelton1991crystal,sosso2016crystal} in which the free-energy barrier for forming a critical nucleus $\Delta G^\ddagger_\mathrm{nuc}\propto \sigma^3/\Delta\mu^2$ where $\sigma$ is the solid--fluid surface free energy and $\Delta\mu$ is the difference in chemical potential between the solid and the fluid. Thus, a lower surface free energy may allow a metastable phase to nucleate more rapidly even if it is metastable.

For the OPP potential, there is good reason to believe that the surface free energies can vary significantly with structure. As seen in the analysis of the contribution of pair distances to the total potential energy in Sect.\ \ref{sect:effect_k} and \ref{sect:effect_phi}, the most stable crystal phases are energetically stabilized by the longer-range interactions associated with the second and third minima. However, the particles at the surface of a cluster will be missing many of those longer-range interactions compared to a particle in bulk, and so the hP2 and cP8 crystals are likely to have the highest surface free energies. 

Furthermore, where possible, the particles at the surface might adopt a somewhat perturbed structure in order to better optimize the energy locally \cite{Doyle25}. This is particularly likely to be the case when the first minimum has a positive energy, as is the case for L1, L2, L3 and I1. Thus, for the L2 IQC cluster illustrated in Fig.\ \ref{fig:IQC_structure}(a) we see a thick surface layer that is depleted in nearest-neighbour bonds. By contrast, for the H3 IQC cluster this effect is much diminished. Although the crystals also show somewhat similar effects, again it seems to be to a lesser extent, perhaps again reflecting the greater structural adaptability of the IQCs. 

Thus, we suggest that the higher surface energies of the most stable crystal phases are likely to be the reason that the IQCs (and BC8 in the low-density region) are in practice observed. Note that significant supercoolings are generally required to observe spontaneous nucleation in simulations and that as the temperature approaches the melting points of the metastable phases, the ordering of the nucleation barriers $\Delta G^\ddagger$ must reverse to favour formation of the most stable phase.

\section{Discussion}

Here, we have shown that the direct coexistence approach provides a useful way to assess the thermodynamic stability of three-dimensional QCs. By contrast, approaches based on thermodynamic integration struggle due to their inability to fully capture the configurational entropy of the IQC.
We note that the current direct coexistence approach becomes more challenging at low coexistence pressures and at temperatures close to a melting-point maximum. However, Gibbs-Duhem integration provides a convenient approach to extend coexistence lines in these regions.

Although the coexistence pressures of the IQCs are expected to be somewhat overestimated due to the use of periodic boundary conditions, the measured coexistence pressures are much higher than for the most stable crystal, enabling us to conclude that the IQCs are never thermodynamically stable for the ten $(k,\phi)$ points in the OPP parameter space that we considered. Hence, the IQC is likely to be only metastable for the complete region of OPP parameter space for which IQCs have been observed to assemble. 
The main reason for this metastability is simply that the IQC exhibits a significantly higher energy than the most stable periodic solids, which cannot be compensated by a higher entropic contribution. 

These findings 
raise the interesting question of whether it is possible to define an isotropic oscillatory potential for which an IQC is thermodynamically stable. 
This study certainly does not exclude that possibility, but the IQC would need to be sufficiently energetically competitive to take advantage of its greater entropy. 
For the current OPP potential, $k$ and $\phi$ do not seem to provide sufficient control over the features of the potential to morph it into one for which the IQC is most stable, but this may be possible for a more flexible potential with more parameters \cite{pan_dshemuchadse_23}.

Although we were not able to fully separate the configurational entropy of the IQC from that due to vibrational anharmonicity, it is clear both that the IQC has significant configurational entropy and that its magnitude decreases as the density of the IQC increases. The latter effect is clearly correlated with the increase in local order associated with icosahedral clusters present in the IQC, specifically the increased occupancy of the dodecahedral sites in the rhombic triacontahedral clusters. This finding emphasises that configurational entropy associated with particle-based QCs not only arises from the analogue of random-tiling entropy, but, even in a one-component system, also includes disorder arising from how different sites are occupied. The decreased entropy associated with the higher-density IQCs may also help to explain why the high-density IQCs have been observed to spontaneously transform to quasicrystalline approximants \cite{Engel2015}, whereas the reverse transition has been observed for the low-density IQC \cite{Je23}.

Another intriguing finding is that the entropy of the BC8 crystalline phase can surpass that of the IQC for the ``higher-density'' $(k,\phi)$ models (e.g.\ I4 and H3 in Fig.\ \ref{Fig:thermo-I4-H3}). This is partly due to the decreased configurational entropy of the higher-density IQCs mentioned above, but also emphasises that differences in vibrational entropy can play a significant role in determining the relative thermodynamic stabilities of different solid phases. For example, for the I2 model, although hP2 has the lowest enthalpy, cP4 is actually the most stable phase at the temperatures we consider.

Here we have reported that part of the phase diagram associated with the low-pressure melting lines of the competing solids. Of course, if one were to determine the complete phase diagram for one of these systems, one would expect additional complexities at high pressure. For example, we showed the presence of melting point extrema in the melting lines of all the considered solids for the I2 model. 
In the case of the higher-density most stable solids (hP2 and cP4) whether these occur before meeting the phase boundary associated with an even higher-density crystal (e.g.\ fcc) at a triple point or in a region where these crystals are metastable is not yet clear.

In future work we plan to use the direct coexistence approach to explore the thermodynamic stability of other three-dimensional quasicrystal-forming systems. This task should probably be relatively straightforward for other one-component systems with isotropic interactions (e.g.\ the IQCs in Refs.\ \onlinecite{Damasceno2017,pan_dshemuchadse_23} and the axial QCs in Refs.\ \onlinecite{Damasceno2017,Ryltsev15,Ryltsev17}). However, it is likely to be more challenging for patchy-particle systems (e.g.\ the one-component IQC in Ref.\ \onlinecite{Noya25} and the octagonal QC in Ref.\  \onlinecite{Kowaguchi25}) as the dynamics of particle rearrangements is expected to be signficantly slower than for the systems studied here. Furthermore, for systems of more than one component, equilibration of the direct coexistence simulations is likely to be even more challenging \cite{Alkemade26}, especially if some subsets of the particles form relatively immobile clusters \cite{Noya21}. 

One interesting aspect of studying the thermodynamics of the axial quasicrystal would be the behaviour of the configurational entropy, as it has been suggested that for the axial QCs to be entropically stabilized there needs to be disorder associated with the periodic dimension \cite{Burkov91,Henley00}. Without such disorder the configurational entropy would not be extensive, and so never be able to overcome any energetic deficit with respect to a periodic crystal. Such ``stacking'' disorder has received relatively little attention, and has only been identified for a few systems \cite{Zhou23,Kowaguchi25}.

Our results imply that the solidification dynamics in these OPP systems are dominated by kinetic effects. Our hypothesis is that this is due to large differences in the surface free energies of the competing solids due to the long-range nature of the OPP potential. It would be interesting to investigate this further by explicitly computing the thermodynamics of nucleation of the competing solid phases.


\section*{Supplementary Material}
The Supplementary Material provides further details 
on (i) the variation of the potential with the model parameters, (ii) snapshots of the initial configurations used in the direct coexistence simulations, (iii) finite size effects, (iv) the structure of the high-density approximant crystal, (v) additional tabulated data related to Figures 9 and 10, and (vi) the order parameters used to distinguish between fluid and solid phases.

\begin{acknowledgments}
CADG and EGN acknowledge the financial support through grant PID2023-151751NB-I00 funded by MCIN/AEI/10.13039/501100011033. C.A.D. acknowledges the predoctoral FPU fellowship 
from Ministerio de Ciencia, Innovación y Universidades (MICIU),
under Grant No. FPU22/00869.
Computer resources and technical assistance provided by CESGA (Centro de Supercomputación de Galicia) and by DRAGO (CSIC) are also gratefully acknowledged.

\end{acknowledgments}

\providecommand{\noopsort}[1]{}\providecommand{\singleletter}[1]{#1}%

\end{document}



\title{Supplementary Material for ``The metastability of icosahedral quasicrystals assembled from one-component systems with isotropic interactions''}

\author{Cosmin A.\ Dicu-Gohoreanu}
\affiliation{Instituto de Qu\'{i}mica F\'{i}sica Blas Cabrera, Consejo Superior de Investigaciones Cient\'{i}ficas, CSIC, Calle Serrano 119, 28006 Madrid, Spain}
\affiliation{Departamento de Química Física I (Unidad Asociada de I+D+i al CSIC), Facultad de  Ciencias Químicas, Universidad Complutense de Madrid, Madrid 28040, Spain}

\author{Jonathan P.~K.\ Doye}
\affiliation{Physical and Theoretical Chemistry Laboratory, Department of Chemistry, University of Oxford, South Parks Road, Oxford, OX1 3QZ, United Kingdom}

\author{Eva G.\ Noya}
\affiliation{Instituto de Qu\'{i}mica F\'{i}sica Blas Cabrera, Consejo Superior de Investigaciones Cient\'{i}ficas, CSIC, Calle Serrano 119, 28006 Madrid, Spain}

\date{\today}

\maketitle

\setcounter{figure}{0}
 \makeatletter
 \renewcommand{\thefigure}{S\@arabic\c@figure}
 \setcounter{equation}{0}
 \renewcommand{\theequation}{S\@arabic\c@equation}
 \setcounter{table}{0}
 \renewcommand{\thetable}{S\@arabic\c@table}
 \setcounter{section}{0}
 \renewcommand{\thesection}{S\@arabic\c@section}

\section{Oscillatory pair potential: variation with $k$ and $\phi$}

\begin{figure*}
\includegraphics[width=0.31\linewidth,clip]{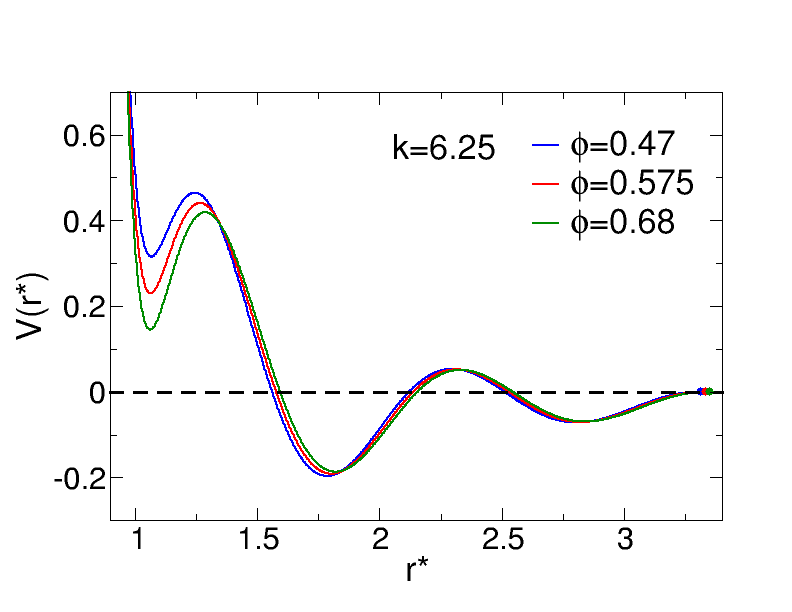} 
\includegraphics[width=0.31\linewidth,clip]{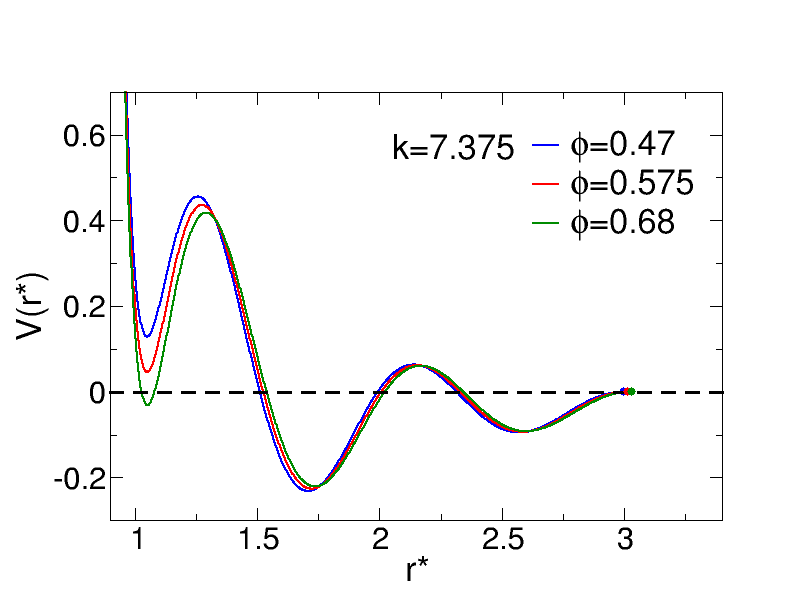} 
\includegraphics[width=0.31\linewidth,clip]{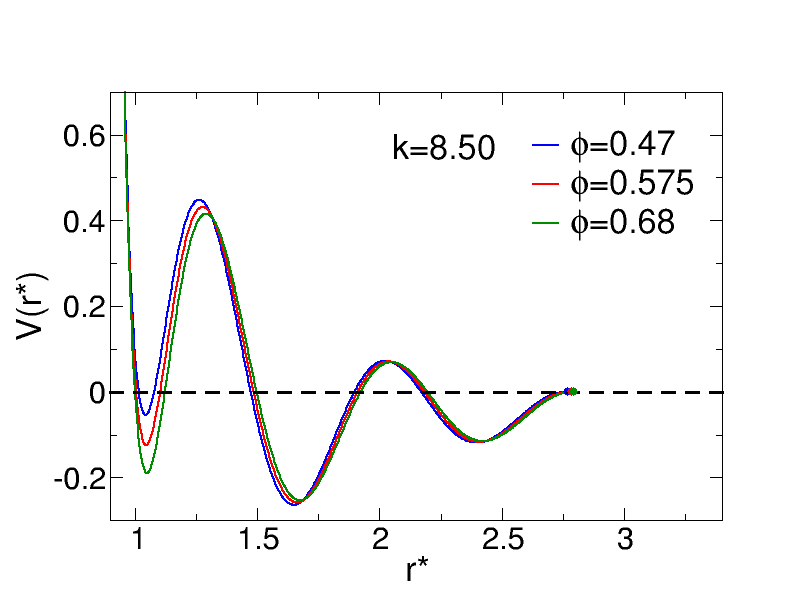} 
\includegraphics[width=0.31\linewidth,clip]{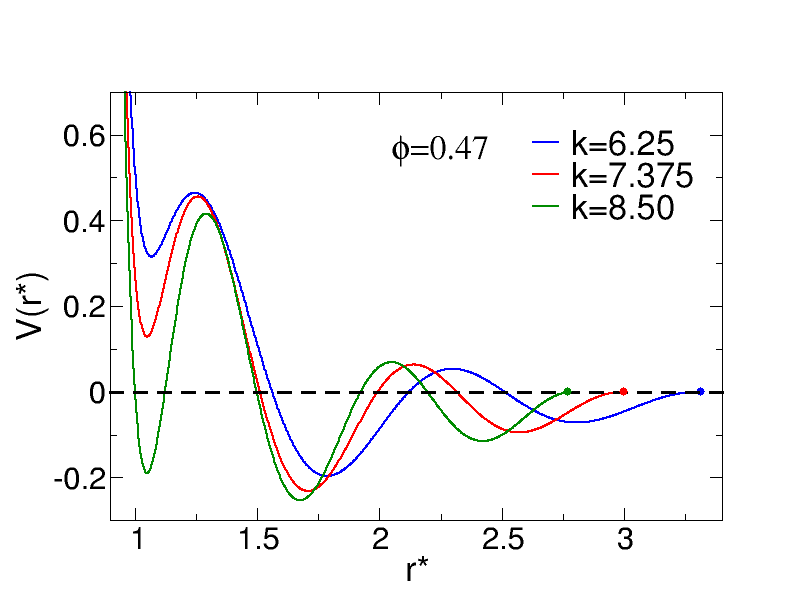} 
\includegraphics[width=0.31\linewidth,clip]{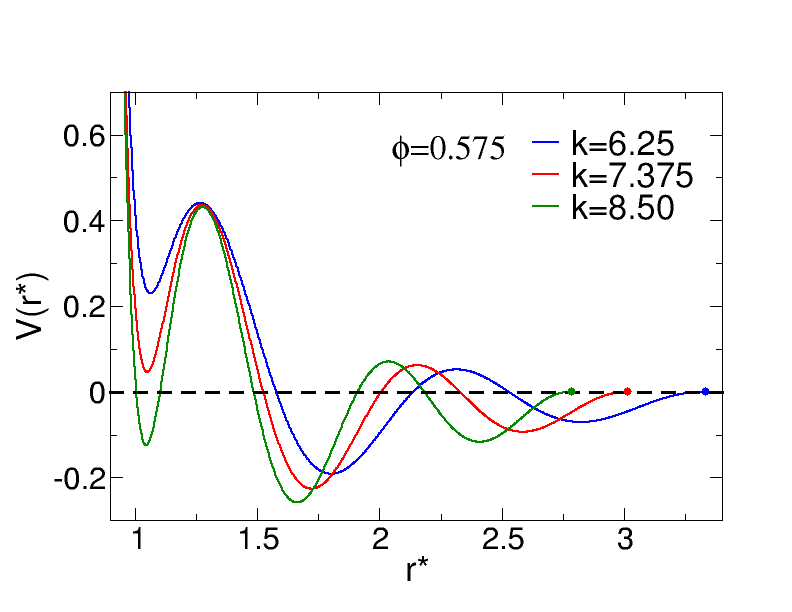} 
\includegraphics[width=0.31\linewidth,clip]{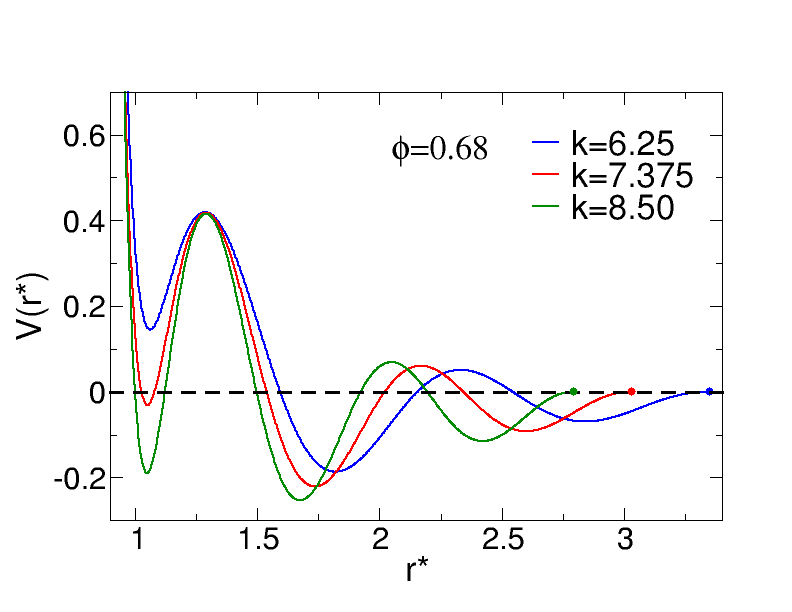} 
\caption{Variation of the potential function across the parameter space explored in this work. In the top row, each graph  displays the change in the potential as $\phi$ increases for several values of $k$. In the bottom row, the evolution of the potential as $k$ increases for three values of $\phi$ is shown.
}
\label{Fig:pot_k_phi} 
\end{figure*}

The potential function  for several combinations of $k$ and $\phi$ spanning the parameter space investigated in this work is shown in Fig.~\ref{Fig:pot_k_phi}.
Keeping $k$ constant, an increase in the phase shift $\phi$ leads to a deeper first minimum and a lower first maximum, the latter also being displaced towards larger distances. The remaining extrema of the potential are less affected, showing only a slight shift to longer distances and being mildly damped. 

On the other hand, when $\phi$ is kept fixed, an increase in the wave vector $k$ leads to more rapid oscillations, significantly reducing the interaction range. The second and third minima and maxima are significantly shifted to shorter distances and become more pronounced (minima get deeper and maxima slightly higher). The first minimum is significantly deepened, becoming negative for the largest $k$ (depending on the value of $\phi$). The first maximum is much less affected, maintaining roughly the same height and position, except for very low values of $\phi$ where it occurs at lower energies and longer distances as $k$ increases. 

\section{Solid-fluid interfaces}
\begin{figure}
\includegraphics[width=90mm]{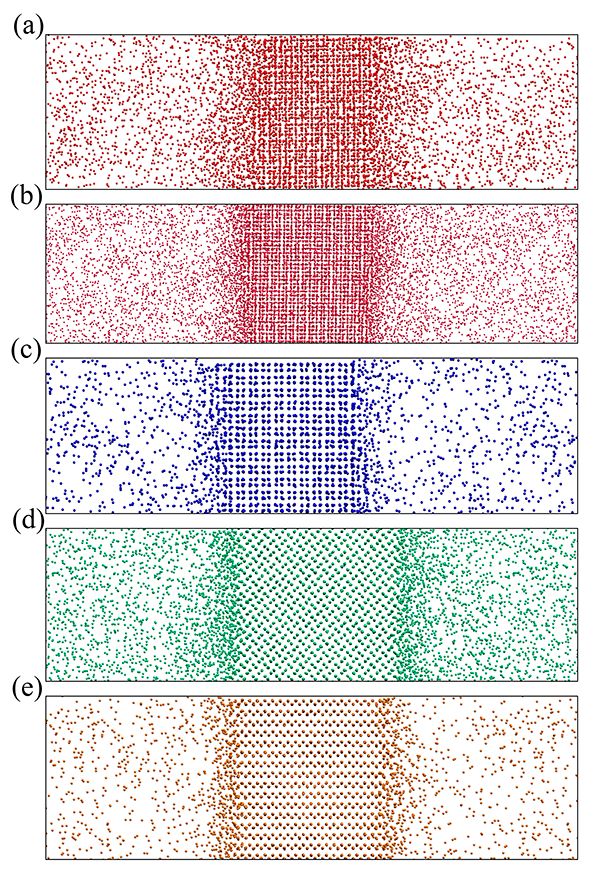} 
\caption{
\label{fig:interfaces}
Snapshots of some of the interfaces simulated in this work: (a) fluid--IQC, (b) fluid--IQC-id, (c) fluid--BC8, (d) fluid--cP4 and (e) fluid--hP2 interfaces. These interfaces were simulated using the I1 model at $T^*=$0.30.}
\end{figure}

Illustrative snapshots from direct coexistence simulations are shown in Fig.\ \ref{fig:interfaces} for the solid structures considered for the I1 model.
Note that the order of stability is evident from the density of particles in the fluid. Also, in all cases a region at the interface is evident that has higher density than the fluid but does not have the same structure as the bulk of the solids.

\section{Finite-size effects in fluid-IQC coexistence}

To assess the effects of the finite size of the periodic boundary conditions on the direct coexistence results for the IQC, different lateral box dimensions were considered. In addition to configurations generated from assembled quasicrystals, we considered both ideal quasicrystals produced by the cut-and-project approach and a series of approximant crystals (ACs) \cite{Noya21}. Note, this approach does not provide a means to straightforwardly vary the density of the ideal IQC or the approximants. Some of their properties are compared in Table \ref{tbl:IQC_AC_I1} for the I1 model where the densities are not too different from the assembled quasicrystals. As noted in the main text, one mechanism for the assembled quasicrystals to vary their density is by partial occupancy of sites within the icosahedral clusters, but varying the average distance between clusters may also play a role. The coexistence pressures of the assembled IQCs and the ACs as a function of box size are compared in Fig.\ 2 in the main text.

\begin{table}
\centering
\caption{\label{tbl:IQC_AC_I1} Average energy and density of the IQC and approximant crystals for the I1 model at $T^*=$0.30 and $p^*=$0.05. IQC-id stands for ideal quasicrystal and IQC-as from a quasicrystal assembled from the fluid phase. The ideal IQC and approximants are those defined in Ref.~\cite{Noya21}. Curiously, the assembled IQC has a slightly lower energy and slightly better packing than the ideal IQC. 
}
\begin{tabular}{lccc} 
\toprule
Phase & $L^*$ & $\rho^*$ & $U^*/N$  \\
\midrule
3/2 AC    &    7.74    &    0.616   &  $-2.19$   \\
3/2 AC ($2\times 2\times 2$) & 15.49 &    0.615   &  -2.20  \\
5/3 AC    &   12.54    &    0.591   &  $-2.11$   \\
8/5 AC    &   20.29    &    0.590   &  $-2.10$   \\
13/8 AC   &   32.83    &    0.590   &  $-2.11$   \\
21/13 AC  &   53.16    &    0.589   &  $-2.12$   \\
IQC-id    &    28.9    &    0.589   &  $-2.09$    \\
IQC-id    &    41.5    &    0.589   &  $-2.09$    \\
IQC-as    &    22.74   &    0.609   &  $-2.15$    \\
IQC-as    &    25.07   &    0.627   &  $-2.21$ \\
IQC-as    &    30.44   &    0.603   &  $-2.11$ \\
IQC-as    &    35.34   &    0.606   &  $-2.11$ \\
IQC-as    &    41.63   &    0.625   &  $-2.19$    \\
IQC-as unor  & 18.21   &    0.668   &  $-2.24$ \\
IQC-as unor  & 22.87   &    0.619   &  $-2.18$ \\
\bottomrule
\end{tabular}
\end{table}

\section{Approximant crystal}
\begin{figure}
\includegraphics[width=70mm]{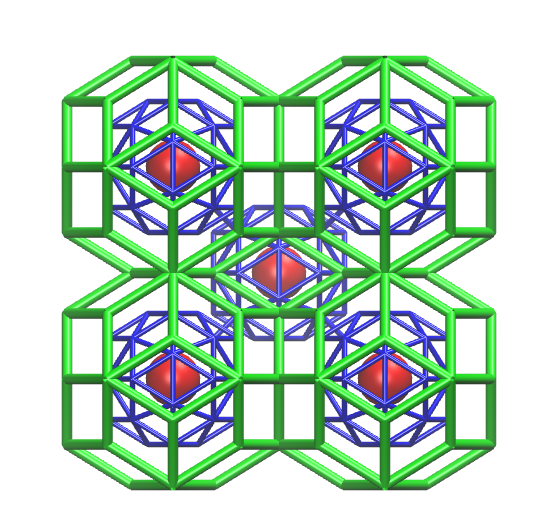} 
\caption{Unit cell of the 1/1 approximant crystal with 78 particles per unit cell reported by Engel et al.}
\label{Fig:AC11_Engel} 
\end{figure}

Engel \emph{et al.} found that in the high-density region of the potential parameter space
the IQC spontaneously transforms into
a 1/1 AC when annealed over a long period \cite{Engel2015}. The 1/1 AC contains 78 particles per unit cell and is formed by a body-centred arrangement of a two-shell cluster formed by two concentric triacontrahedra. It is illustrated in Fig.\ \ref{Fig:AC11_Engel}. Note, it has a higher density than the approximants discussed in the previous section. As can be seen in Fig.\ 6 in the main text, our direct coexistence simulations predict similar stabilities for both phases within the statistical uncertainty of this approach.

\section{Effect of the model parameters on the $p$--$T$ phase diagram}

Tables \ref{tbl:ener_L2_L3_I2} and \ref{tbl:neigh_L2_I2} compare the properties of the solids at coexistence with the fluid for the L2 and I2 models, which just differ by the value of $k$.

\begin{table*}
\centering
\caption{\label{tbl:ener_L2_L3_I2} Thermodynamic properties at coexistence at $T^*$=0.27 for potentials with $\phi=0.5$ and increasing $k$. }
\begin{tabular}{ccccccccccccccccccccc}
\toprule
Model  & & Solid & &$p_\mathrm{coex}$ & & $u_\mathrm{f}$ & & $u_\mathrm{s}$ & & $\Delta u_\mathrm{sf}$ & & $\rho_\mathrm{f}$ & & $\rho_\mathrm{s}$  & & $h_\mathrm{f}$ & & $h_\mathrm{s}$ & & $\Delta h_\mathrm{sf}$ \\
\midrule
L2 & & BC8 & & 0.016 & & $-0.43$ & & $-2.05$ & & $-1.62$ & & 0.140 & & 0.572  & & $-0.32$ & & $-2.02$ & & $-1.71$ \\
L2 & & hP2 & & 0.0096 & & $-0.17$ & & $-2.62$ & & $-2.45$ & & 0.050 & & 0.902 & &  0.02 & & $-2.61$ & & $-2.63$ \\
L2 & & cP4 & & 0.118 & & $-0.99$ & & $-1.92$ & & $-0.93$ & & 0.411 & & 0.676  & & $-0.70$ & & $-1.75$ & & $-1.05$ \\
L2 & & IQC & & 0.120 & & $-0.99$ & & $-1.78$ & & $-0.79$ & & 0.415 & & 0.552  & & $-0.70$ & & $-1.56$ & & $-0.86$ \\
\midrule
I2 & & BC8 & & 0.0021 & & $-0.03$ & & $-2.74$ & & $-2.71$ & & 0.008 & & 0.648 & &  0.232  & & $-2.74$ & & $-2.96$  \\
I2 & & hP2 & & 0.0028 & & $-0.04$ & & $-3.05$ & & $-3.01$ & & 0.011 & & 0.965 & &  0.205  & & $-3.05$ & & $-3.26$  \\
I2 & & cP4 & & 0.0012 & & $-0.02$ & & $-2.99$ & & $-2.97$ & & 0.005 & & 0.763 & &  0.220  & & $-2.99$ & & $-3.21$  \\
I2 & & IQC & & 0.0032 & & $-0.05$ & & $-2.64$ & & $-2.58$ & & 0.013 & & 0.728 & &  0.196  & & $-2.64$ & & $-2.83$  \\
\bottomrule
\end{tabular}
\end{table*}

\begin{table*}
\centering
\caption{\label{tbl:neigh_L2_I2} Average number of neighbours for the considered solid phases for the ranges of the potential illustrated in Fig.\ 9(g) (L2) and (h) (I2).}
\begin{tabular}{lcccccc} 
\toprule
& \multicolumn{6}{c}{L2 potential - distance range ($\Delta r$)} \\ 
\cmidrule(lr){2-7}
Phase & 1.0--1.26 & 1.26--1.52 & 1.52--2.03$^\text{N}$ & 2.03--2.38$^\text{P}$ & 2.38--3.10$^\text{N}$ & $< 3.10$ \\
\midrule
BC8  & 3.9       & 1.1        & 17.8       & 3.4        & 46.1       & 72.3 \\
hP2  & 8.0       & 0.8        & 29.9       & 5.2        & 65.4       & 109.5 \\
cP4  & 5.7       & 0.5        & 18.8       & 7.0        & 50.5       & 82.6 \\
IQC  & 3.9       & 0.9        & 17.1       & 4.0        & 43.4       & 69.3 \\
\midrule
& \multicolumn{6}{c}{I2 potential - distance range ($\Delta r$)} \\ 
\cmidrule(lr){2-7}
Phase & 1.0--1.27 & 1.27--1.49 & 1.49--1.94$^\text{N}$ & 1.94--2.23$^\text{P}$ & 2.23--2.87$^\text{N}$ & $< 2.87$ \\
\midrule
BC8  & 4.0       & 1.2        & 17.7       & 2.6        & 41.3       & 66.9 \\
hP2  & 8.0       & 0.8        & 28.9       & 2.3        & 58.6       & 98.7 \\
cP4  & 6.0       & 0.1        & 18.2       & 5.8        & 43.5       & 73.6 \\
IQC  & 5.7       & 0.8        & 18.1       & 5.4        & 43.0       & 73.1 \\
\bottomrule
\multicolumn{7}{l}{\small $^\text{P}$ Positive potential ($u_\mathrm{OPP}(r)>0$); $^\text{N}$ Negative potential ($u_\mathrm{OPP}(r) \leq 0$).}
\end{tabular}
\end{table*}

Tables \ref{tbl:ener_I4_H1_H3} and \ref{tbl:neigh_I4_H3} compare the properties of the solids at coexistence with the fluid for the I4 and H3 models, which just differ by the value of $\phi$.

\begin{table*}
\centering
\caption{\label{tbl:ener_I4_H1_H3} Thermodynamic properties at coexistence at $T^*$=0.33 for potentials with $k=8.5$ and increasing $\phi$.}
\begin{tabular}{ccccccccccccccccccccc}
\toprule
Model  & & Solid & &$p_\mathrm{coex}$ & & $u_\mathrm{f}$ & & $u_\mathrm{s}$ & & $\Delta u_\mathrm{sf}$ & & $\rho_\mathrm{f}$ & & $\rho_\mathrm{s}$  & & $h_\mathrm{f}$ & & $h_\mathrm{s}$ & & $\Delta h_\mathrm{sf}$ \\
\midrule
I4 & & BC8 & & 0.020 & & $-0.12$ & & $-2.75$ & & $-2.63$ & & 0.103 & & 0.676 & &  0.07& & $-2.72$ & & $-2.79$ \\
I4 & & cP4 & & 0.006 & & $-0.07$ & & $-3.23$ & & $-3.16$ & & 0.021 & & 0.794 & &  0.23 & & $-3.21$ & & $-3.44$ \\
I4 & & IQC & & 0.030 & & $-0.79$ & & $-2.74$ & & $-1.95$ & & 0.279  & & 0.718 & & $-0.68$ & & $-2.69$ & & $-2.01$ \\
\midrule
H3 & & BC8 & & 0.0100 & & $-0.12$ & & $-2.88$ & & $-2.76$ & & 0.035 & & 0.656  & & 0.166  & & $-2.86$ & & $-3.03$ \\
H3 & & hP2 & & 0.0089 & & $-0.09$ & & $-3.39$ & & $-3.30$ & & 0.026 & &  0.962 & & 0.252  & & $-3.38$ & & $-3.63$ \\
H3 & & cP4 & & 0.0026 & & $-0.03$ & & $-3.47$ & & $-3.44$ & & 0.008 & & 0.771  & & 0.295  & & $-3.47$ & & $-3.76$ \\
H3 & & IQC & & 0.0065 & & $-0.07$ & & $-3.11$ & & $-3.03$ & & 0.021 & &  0.769 & & 0.240  & & $-3.10$ & & $-3.34$ \\
\bottomrule
\end{tabular}
\end{table*}

\begin{table*}
\centering
\caption{\label{tbl:neigh_I4_H3} Average number of neighbours for the considered solid phases for the ranges of the potential illustrated in Fig.\ 10(g) (I4) and (h) (H3).}
\begin{tabular}{lcccccc} 
\toprule
& \multicolumn{6}{c}{I4 potential - distance range ($\Delta r$)} \\ 
\cmidrule(lr){2-7}
Phase & 1.0--1.26  & 1.26--1.48  & 1.48--1.90$^\text{N}$ & 1.90--2.17$^\text{P}$ & 2.17--2.77$^\text{N}$ & $< 2.77$ \\
\midrule
BC8  &    4.0    &     1.6    &   17.2     &    2.2     &    38.2    & 56.5  \\
hP2  &    8.0    &     0.9    &   27.4     &    3.7     &    48.5    & 94.9 \\
cP4  &    6.0    &     0.2    &   17.9     &    5.9     &    41.9    & 73.8  \\
IQC  &    5.2    &     1.1    &   17.3     &    4.0     &    36.7    & 67.6  \\
\midrule
& \multicolumn{6}{c}{H3 potential - distance range ($\Delta r$)} \\ 
\cmidrule(lr){2-7}
Phase & 1.0--1.29 & 1.29--1.50 & 1.50--1.92$^\text{N}$ & 1.92--2.20$^\text{P}$ & 2.20--2.79$^\text{N}$ & $< 2.79$ \\
\midrule
BC8  &    4.1    &    1.6     &    17.1    &     2.4    &    37.3    & 55.4 \\
hP2  &    8.0    &    1.1     &    27.6    &     3.3    &    48.2    & 92.2 \\
cP4  &    6.0    &    0.3     &    17.9    &     5.8    &    41.7    & 71.5 \\
IQC  &    6.2    &    1.0     &    17.7    &     6.6    &    38.7    & 72.7 \\
\bottomrule
\multicolumn{7}{l}{\small $^\text{P}$ Positive potential ($u_\mathrm{OPP}(r)>0$); $^\text{N}$ Negative potential ($u_\mathrm{OPP}(r) \leq 0$).}
\end{tabular}
\end{table*}

\section{Order parameters for distinguishing local environments}

\begin{figure*}
\includegraphics[width=\linewidth,clip]{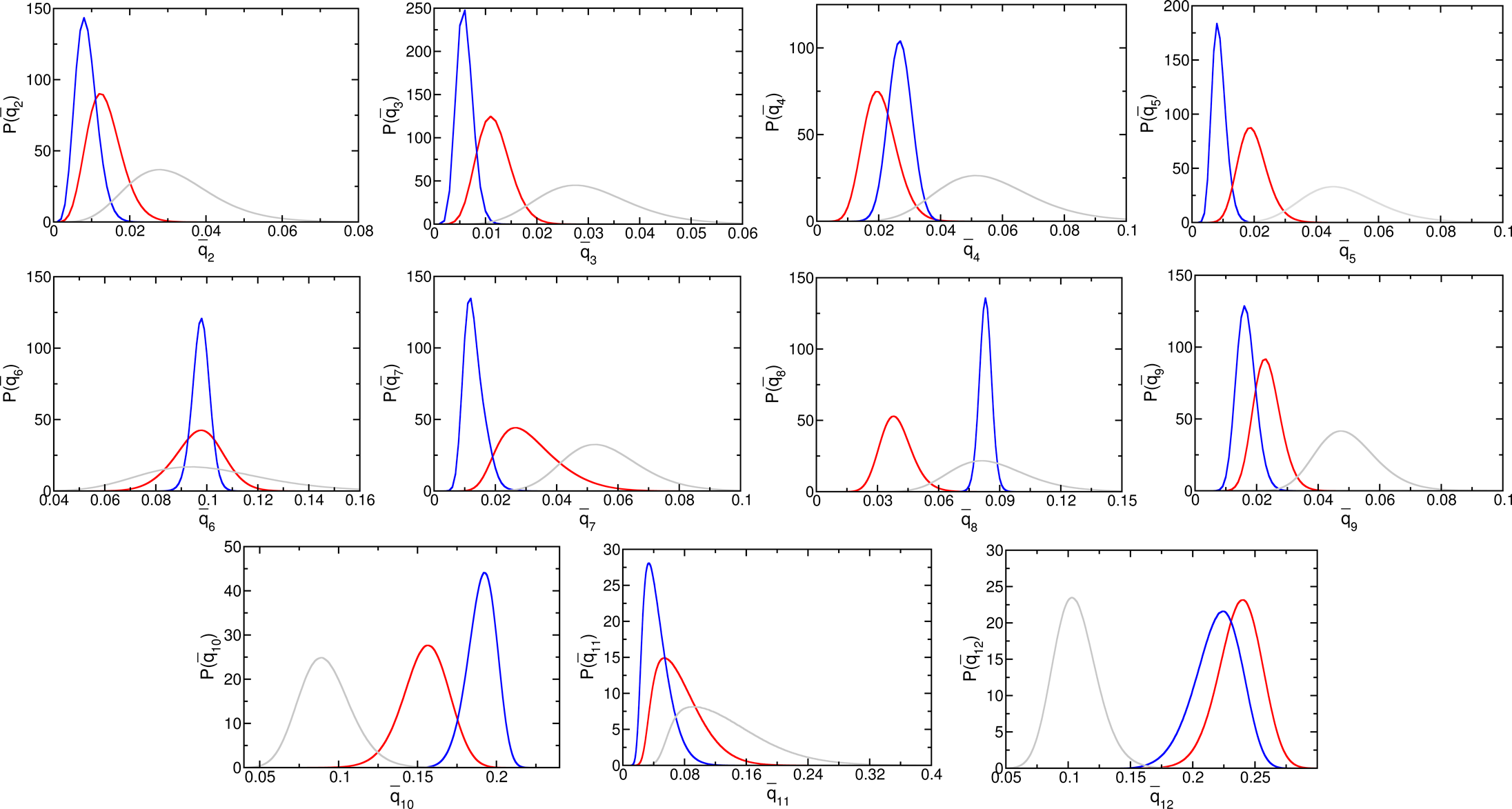}
\caption{\label{fig:OP_IQCvBC8}
Neighbour-averaged local bond order parameters $\bar{q}_i$ ($i$=2--12) \cite{Dellago2008} for IQC (red), BC8 (blue) and fluid (grey) for the L2 model at $T$=0.27 and the coexistence pressure. Particles are considered neighbours if their interparticle distance is shorter than 2.01, approximately corresponding to the second minimum in the radial distribution function of each phase.} 
\end{figure*}

\begin{figure*}
\includegraphics[width=\linewidth,clip]{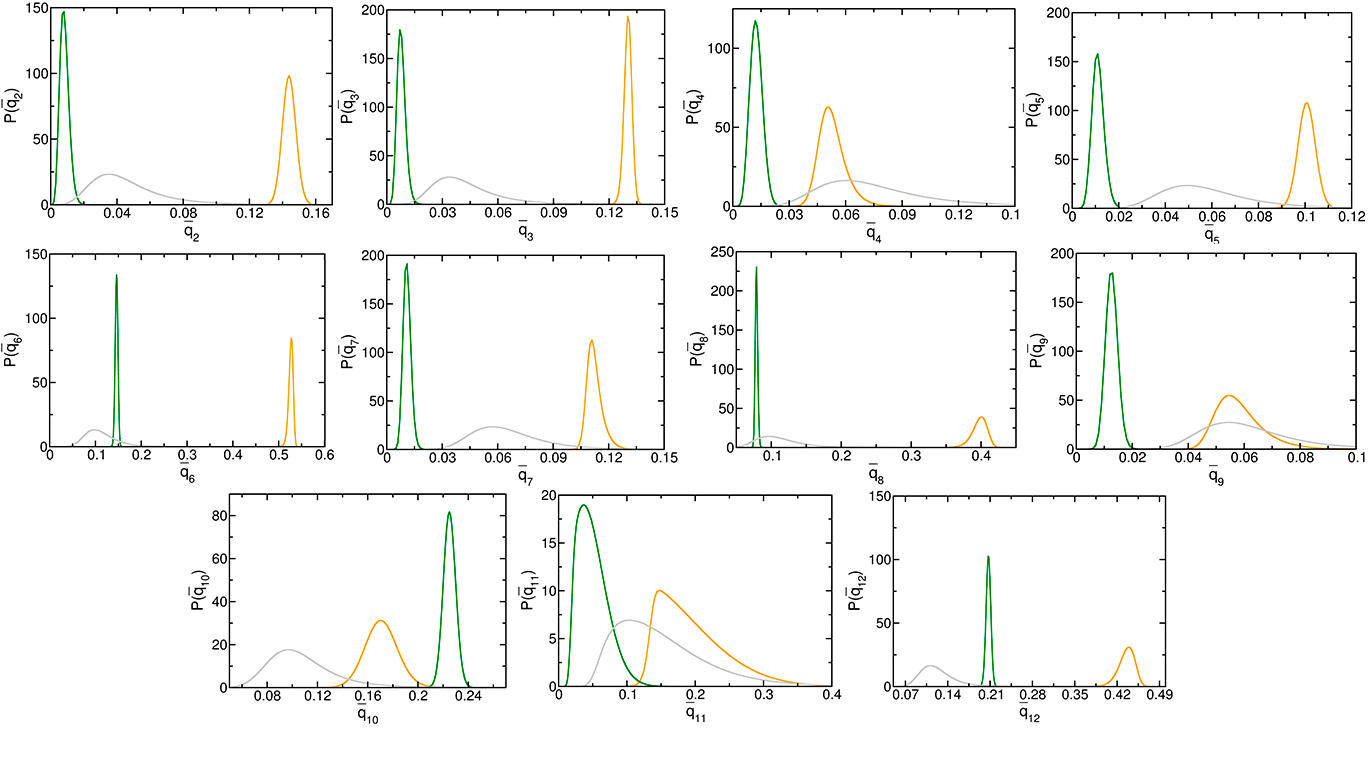}
\caption{\label{fig:OP_cP4vhP2}
Neighbour-averaged local bond order parameters $\bar{q}_i$ ($i$=2--12) \cite{Dellago2008} for cP4 (green), hP2 (orange) and fluid (grey) for I4 at $T^*$=0.30 and the coexistence pressure. Particles are considered neighbours if their inter-particle distance is shorter than 2.01, corresponding to the second minimum in the radial distribution function of each phase.} 
\end{figure*}

\begin{figure*}
\includegraphics[width=0.7\linewidth,clip]{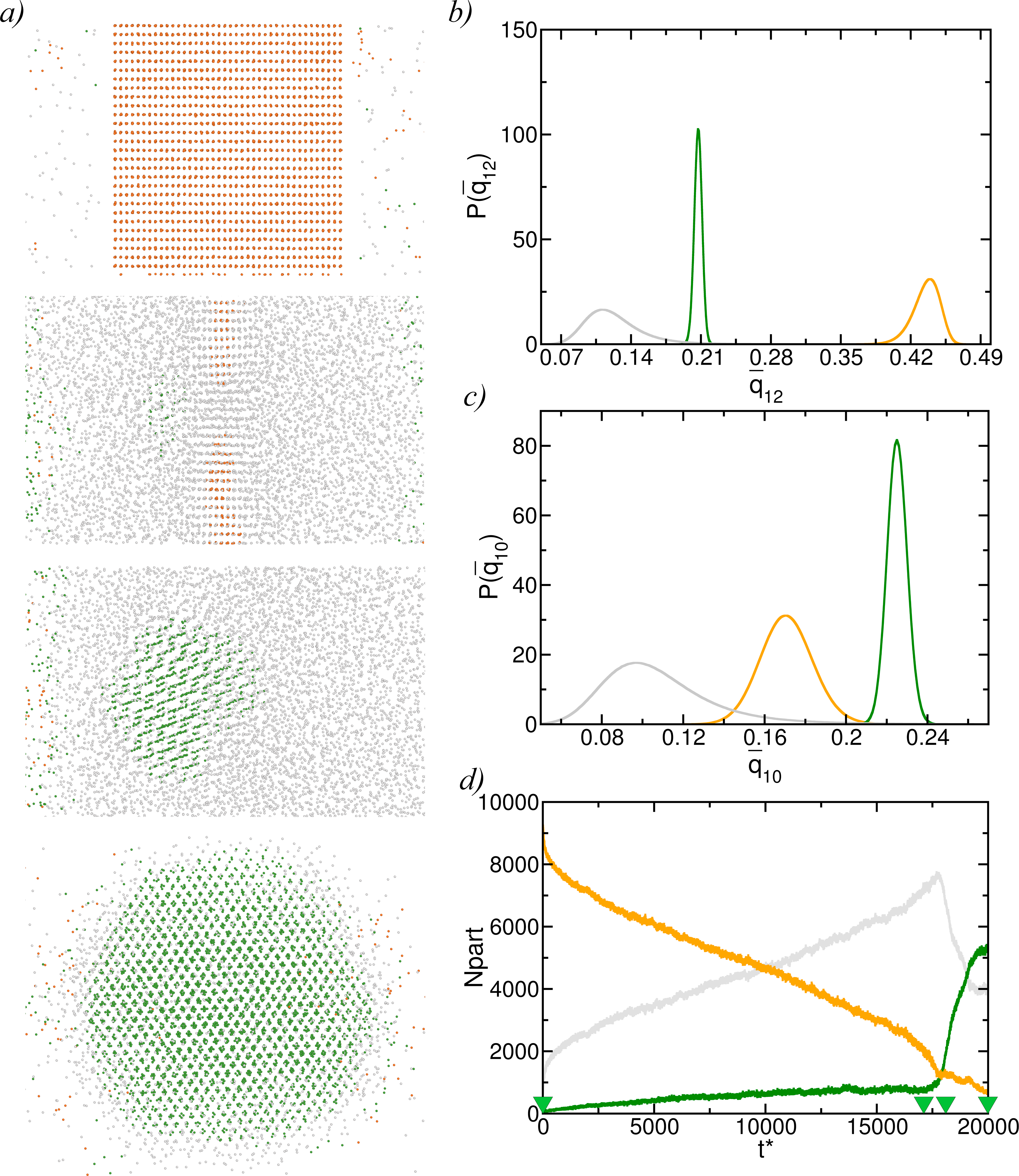}
\caption{\label{fig:hP2tocP4}
Nucleation and growth of cP4 at the fluid-hP2 interface for the I4 model at $T^*$=0.30. (a) Snapshots at different times show the initial fluid–hP2 interface ($t^*=0$), the melting of hP2 ($t^*=17\,200$), the nucleation of cP4 in the bulk of the fluid ($t^*=18\,000$), and the final state ($t^*=20\,000$) in which the fluid coexists with cP4. (b) Particles in hP2 local environments are distinguished from the rest using $\bar{q}_{12}$ with a threshold of $\bar{q}_{12} = 0.35$ being used. 
(c) The remaining particles are subsequently classified as cP4 or fluid using $\bar{q}_{10}$ with a threshold of $\bar{q}_{10} = 0.19$ being used. Both  $\bar{q}_{10}$ and $\bar{q}_{12}$ were calculated using a cutoff distance of 2.01 to define neighbours. (d) Evolution of the number of particles in fluid, hP2, and cP4 environments as a function of time. 
} 
\end{figure*}

In cases where transitions between solid phases were observed in the direct coexistence simulations, we used neighbour-averaged local bond-order parameters \cite{Dellago2008} to distinguish particles belonging to the different phases and so better track the transitions. Figs.\ \ref{fig:OP_IQCvBC8} and \ref{fig:OP_cP4vhP2} show the $\bar{q}_i$ distributions for $i$=2--12 for the relevant phases. These formed the basis for the choices used in Fig.\ 7 in the main text and in Fig.\ \ref{fig:hP2tocP4} to follow the transitions from IQC to BC8 and hP2 to cP4, respectively.

\providecommand{\noopsort}[1]{}\providecommand{\singleletter}[1]{#1}%